\documentclass[12pt,preprint]{aastex}
\newcommand{\simle}{\mbox{$\stackrel{<}{_{\sim}}$}}
\newcommand{\simge}{\mbox{$\stackrel{>}{_{\sim}}$}}
\newcommand\arcdegg{\hbox{$^\circ$}}

\newcommand\msun{\hbox{\,M$_\odot$}}

\newcommand\lsun{\hbox{\,L$_\odot$}}

\newcommand\vis{${\mathcal{V}^2}$}
\newcommand\viss{${\mathcal{V}^2}$ }

\shorttitle{YSO Skew IOTA3}
\shortauthors{Monnier et al.}

\begin{document}


\title{Few Skewed Disks Found in First Closure-Phase Survey of Herbig Ae/Be stars}


\author{J.~D.~Monnier\altaffilmark{1}, J.-P.~Berger\altaffilmark{2},
R.~Millan-Gabet\altaffilmark{3}, W.~A.~Traub\altaffilmark{4,5},
F.~P.~Schloerb\altaffilmark{6}, E.~Pedretti\altaffilmark{1},
M.~Benisty\altaffilmark{2}, N.~P.~Carleton\altaffilmark{4},
P.~Haguenauer\altaffilmark{7}, P.~Kern
\altaffilmark{2}, P.~Labeye\altaffilmark{8},
M.~G.~Lacasse\altaffilmark{4}, F.~Malbet\altaffilmark{2},
K.~Perraut\altaffilmark{2}, M.~Pearlman\altaffilmark{4},
 and M.~Zhao\altaffilmark{1} }

\altaffiltext{1}{monnier@umich.edu: University of Michigan Astronomy Department, 941 Dennison Bldg, Ann Arbor, MI 48109-1090, USA.}
\altaffiltext{2}{Laboratoire d'Astrophysique de Grenoble, 414 Rue de la Piscine 38400 Saint Martin d'Heres, France}
\altaffiltext{3}{Michelson Science Center, California Institute of Technology, Pasadena, CA}
\altaffiltext{4}{Harvard-Smithsonian Center for Astrophysics, 60 Garden St,
Cambridge, MA, 02138, USA}
\altaffiltext{5}{Jet Propulsion Laboratory, California Institute of Technology, Pasadena, CA}
\altaffiltext{6}{University of Massachusetts, Amherst}
\altaffiltext{7}{European Southern Observatory}
\altaffiltext{8}{LETI-CEA, Grenoble, France}


\begin{abstract}
  Using the 3-telescope IOTA interferometer on Mt. Hopkins, we report
  results from the first near-infrared ($\lambda=1.65\mu$m)
  closure-phase survey of Young Stellar Objects (YSOs).  These closure
  phases allow us to unambiguously detect departures from
  centrosymmetry (i.e., skew) in the emission pattern from YSO disks
  on the scale of $\sim$4~milliarcseconds, expected from generic
  ``flared disk'' models.  Six of fourteen targets showed small, yet
  statistically-significant, non-zero closure phases, with largest
  values from the young binary system MWC~361-A and the (pre-main
  sequence?) Be star HD~45677.  Our observations are quite sensitive
  to the {\em vertical} structure of the inner disk and we confront
  the predictions of the ``puffed-up inner wall'' models of Dullemond,
  Dominik, and Natta (DDN).  Our data support disks models with {\em
    curved} inner rims because the expected emission appear
  symmetrically-distributed around the star over a wide range of
  inclination angles. In contrast, our results are {\em incompatible}
  with the models possessing {\em vertical} inner walls because they
  predict extreme skewness (i.e., large closure phases) from the
  near-IR disk emission that is not seen in our data.  In addition, we
  also present the discovery of mysterious H-band ``halos''
  ($\sim$5-10\% of light on scales 0.01-0.50\arcsec) around a few
  objects, a preliminary ``parametric imaging'' study for HD~45677,
  and the first astrometric orbit for the young binary MWC~361-A.

\end{abstract}

\keywords{accretion disks -- instrumentation: interferometers --
  techniques: interferometric -- stars: binaries: spectroscopic --
  stars: pre-main sequence -- stars: individual (AB Aur, HD 45677, HD
  144432, MWC 166, MWC 275, MWC 297, MWC 342, MWC 361, MWC 480, MWC
  614, MWC 863, MWC 1080, RY Tau, v1295 Aql) }


%\tableofcontents

\section{Introduction}
The study of accretion disk evolution in Young Stellar Objects (YSOs)
promises to directly link the fields of star formation with planet
formation.  Spatially-resolved, multi-wavelength observations can
directly probe the dust and gas distribution on solar system scales,
allowing precise testing of disk models while simultaneously providing
the crucial physical parameters for planet formation initial
conditions.  Young Jupiter-mass planets embedded in natal disks should
even be detectable in the future.

Already observations from long-baseline interferometers (IOTA, PTI,
Keck) have empirically determined the near-infrared (NIR) emission
sizes of YSOs,
\citep[e.g.,][]{rmg1999a,rmg2001,akeson2000,eisner2003}, finding the
sizes to be strongly linked to the luminosity of the central stars
\citep{mm2002,monnier2005}.  These results are qualitatively explained
by a model where the NIR emission arises entirely from the hottest
dust at the inner edge of an accretion disk, and the measured sizes
correspond to the dust-destruction radius around the young stars
\citep{lkha2001,natta2001}.  Recently, \citet{eisner2004} reported
convincing signs that NIR dust emission from many intermediate-mass
Herbig Ae/Be stars are elongated, implicating disk-like structure as
expected from star and planet formation theory.

The simple detection of ``elongation'' by an interferometer, however,
does not well-constrain the geometry of the emission region.  For
instance, current data can not distinguish between a) a
centrosymmetric emission pattern as expected from a perfectly flat,
inclined disk, b) elongated emission arising preferentially from one
side of the disk \citep[as expected from optical depth effects
associated with flaring;][]{malbet2001}, or c) hybrid disk+halo models
with multiple emission components \citep{vinkovic2006}.  All current
results are based on measuring interferometer fringe {\em amplitudes},
while interferometric fringe {\em phase information} is required to
unambiguously detect deviations from simple symmetries. Ultimately,
the phases and amplitudes are all needed to permit model-independent
imaging using aperture synthesis techniques \citep{tms2001}.

While atmospheric turbulence corrupts the direct measurement of fringe
phase, interferometrists using three or more telescopes can measure
the ``closure phase'' (CP), a phase quantity that is unspoiled by
telescope-specific phase errors \citep{jennison58,monnier_mss}.
Recently, the Infrared Optical Telescope Array (IOTA) became the third
facility to successfully achieve closure phase operation
\citep{traub2003,monnier2004b} and is the only such facility that has
demonstrated the capability to study Young Stellar Objects
\citep{rmg2006}.

Here, we expand the original Herbig Ae/Be survey work of
\citet{rmg2001} using three IOTA telescopes simultaneously,
tripling the rate at which visibility measurements are collected and
allowing  closure phases to be measured for the first time for a
sample of YSOs \citep[see][for the first results for the prototype
Herbig Ae star AB~Aur]{rmg2006}.  This work has been made possible by
the advanced beam combiner IONIC3 \citep{berger2003} which exploits
integrated optics technology developed for the telecommunication
industry.

This article presents a large volume of data and analysis and is
organized in the following sections.  First, we describe the
interferometer observations and present the data results in a summary
form.  Second, we analyze the spectral energy distribution and
visibility data together for each target in order to estimate the NIR
emission size; this step is crucial for interpreting the closure phase
results.  Sections 4 \& 5 contain the most novel aspects of this
paper. Here we outline our method for using closure phases to probe
the amount of skew in the YSO disk emission.  Then, we apply our
method to quantify the {\em skewed} disk emission and discuss the
implications on current disk models. Lastly, we treat the special case
of HD~45677 and report the first astrometric orbit for the YSO binary
MWC~361-A.

\section{Observations and Data Reduction}
\subsection{Description of IOTA Observations}
\label{section:observations}
All the data presented herein were obtained using the IOTA (Infrared-Optical
Telescope Array) interferometer \citep{traub2003}, a long baseline
interferometer which observes at visible and near-infrared
wavelengths.  Located on Mt. Hopkins (Arizona), IOTA is operated by a consortium of institutions, most
notably the Smithsonian Astrophysical Observatory and the University
of Massachusetts at Amherst.  The three 0.45-m telescopes are movable among
17 stations along 2 orthogonal linear arms (telescopes A \& C can move
along the 35-m northeastern arm, while Telescope B moves along the
15-m southeastern arm).  By observing a target in many different array
configurations, IOTA can synthesize an aperture 35m$\times$15m
(corresponding to an angular resolution of
$\sim$5$\times$12~milliarcseconds at 1.65$\mu$m).
  
First results using closure phases have been published by
\citet{monnier2004b} and \citet{kraus2005} using the IONIC3 combiner
on IOTA.  We only just introduce the IONIC3 combiner here and refer
the interested reader to a recent description in \citet{berger2003};
an engineering paper with detailed description of the optical
component and its performance is in preparation (Berger et al. 2006).
Light from each telescope is focused into a dedicated single-mode
fiber and the three fibers are aligned using a silicon v-groove array
mated to planar waveguides on the integrated optics (IO) device.  The
optical circuit acts to split the light from each fiber/telescope
before recombining each telescope pair (AB, BC, AC) at three IO
couplers.  This ``pair-wise'' combination scheme leads to six
interferometer channels (two for each baseline) and the fringes are
detected using a sensitive HgCdTe (Rockwell PICNIC) detector
\citep{pedretti2004}.  Varying coupling efficiencies into the fibers
(due to seeing variations and tip-tilt control errors) cause the
system visibilities to vary in time, however this effect can be
directly calibrated using measurements of the IONIC3 (optical) flux
transfer matrix which relates the six output channels to the three
input channels for incoherent light \citep[explained further
in][]{monnier2004b}.

The interference fringes are temporally-modulated on the detector by
scanning piezo mirrors placed in two of the three beams of the
interferometer.  A typical single observation consists of 200 scans
obtained in $\sim$4~min, followed by calibration measurements of the
background and single-telescope fluxes (important for characterizing
the IO beamsplitters/couplers).  Target observations are interleaved
with an identical ``calibration'' sequence obtained on an unresolved
or partially-resolved star, serving to calibrate the
interferometer's instrumental response and effects of atmospheric
seeing on the visibility amplitudes.  The target and calibrator
sources are typically separated on the sky by 5-10 degrees and are
observed 10-20 minutes apart; these conditions ensure that the
calibrator observations provide a good estimate of the instrument's
transfer function.

The target sample was selected from the catalog of Herbig Ae/Be
members by \citet{the1994}.  Eleven of the 14 targets come from Table
1 of {\em bona fide} Herbig Ae/Be members or candidate members.
HD~45677 is contained in Table 3 ``Extreme emission line objects'' and
MWC~342 is found in Table 4a ``Other early type emission line stars
with IR excess''; both of these objects are discussed individually in
\S\ref{individual}.  Lastly, we have included one object not listed in
the Th\'{e} catalog, RY~Tau.  RY~Tau (F8) is often considered too
late-type to be a Herbig Ae/Be star and too early to be a T~Tauri
star. However, the inner disks of T~Tauri and Herbig Ae stars seem to
share common physics \citep{rmg_ppv, mm2002} and we so have included
this target in our sample.  We note that the IOTA Interferometer with
the IONIC combiner has a limiting H-band magnitude of $\sim$7, and
this limitation played a major role in our selections.

\begin{figure}[hbt]
\begin{center}
\includegraphics[width=6in,clip]{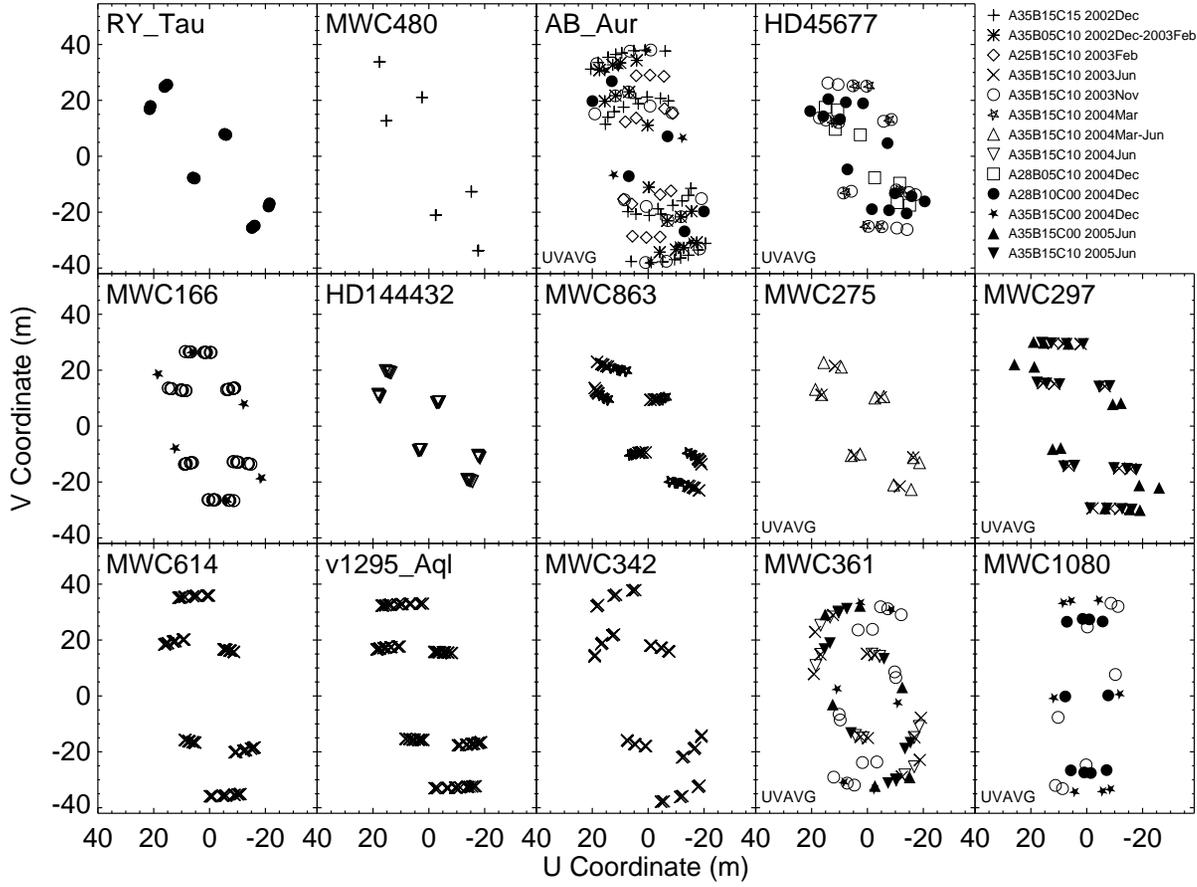}
\figcaption{\footnotesize These panels show the (u,v)-coverage of the
  data for all targets presented in this paper.  Each observing
  configuration and epoch are denoted by a unique plotting symbol
  according to the legend in the upper-right panel.  For a few targets
  with large numbers of data points, we have averaged the data in the
  (u,v)-plane for presentation clarity and these panels are denoted by
  the word ``UVAVG'' in the bottom-left corner.
\label{fig_uvcov}}
\end{center}
\end{figure}

The science targets, along with their fundamental parameters,
are listed in Table~\ref{targets}.  All the science targets presented
here were observed using a standard H band filter ($\lambda_0 =
1.65\mu$m, $\Delta\lambda = 0.30\mu$m).  Full observing information
for our science targets can be found in Tables~\ref{table_obslog} \&
\ref{calibrators}, including dates of observation, interferometer
configurations, and calibrator details; note that the calibrators are
all nearly unresolved on our baselines and the size uncertainties do
not generally significantly contribute to the calibration
uncertainties.  Figure~\ref{fig_uvcov} shows the (u,v)-coverage
obtained during each epoch for each science target.  In the next
section, we discuss our analysis of these data and the techniques used
to process raw data into calibrated squared-visibilities (\vis) and
closure phases (CP).

\subsection{Data Reduction}

Reduction of the IONIC3 visibility data was carried out using custom
software similar in its main principles to that described by
\citet[][for the FLUOR experiment]{foresto1997}, developed using the
{\em Interactive Data Language} (IDL).  In short, we measure the power
spectrum of each interferogram (proportional to the target \vis),
after correcting for intensity fluctuations and subtracting out bias
terms from read noise, residual intensity fluctuations, and photon
noise \cite[e.g.,][]{perrin2003a}.  We require that interferograms are
detected for at least two of the three baselines in order to assure a
good closure phase measurement; this condition is almost always met
thanks to a realtime fringe packet ``tracker'' (discussed further
below).  Lastly, the data pipeline applies a correction for the
variable flux ratios for each baseline by using a flux transfer matrix
\citep[e.g.,][]{foresto1997,monnier2001}.  We have studied our
calibration accuracy by extensive observations of the binary star
$\lambda$~Vir \citep{zhao2006}. For bright stars (H mag $\simle$5), we
have validated 2\% calibration error for \viss (corresponding to 1\%
error in visibility).  However, most YSOs are faint requiring slow
scan rates and exposing our calibration to seeing changes.  Under most
conditions for faint stars, we find a calibration error of $\sim$5\%
in visibility-squared (e.g., 2.5\% in visibility) although errors of
$\simge$10\% are occasionally encountered during poor seeing
conditions. For very resolved objects, the calibration error is
dominated by imperfect subtraction of the power spectrum background
resulting in a $\Delta$\vis$=0.02$ noise floor.  To be conservative,
we adopted a uniform 5\% calibration error (with $\Delta$\vis$=0.02$
noise floor) for all \viss in this paper, although this overestimates
the error for many epochs \citep[][adopted similar values for
AB~Aur]{rmg2006}.

In order to measure the closure phase, a fringe-tracking algorithm was
applied in realtime while recording interferograms
\citep{pedretti2005a}, ensuring that interference occurs nearly
simultaneously for all baselines.  We followed the method of
\citet{baldwin1996} for calculating the complex triple amplitude in
deriving the closure phase, explicitly guaranteeing that the fringe
frequencies for each triple product also ``close'' ($\nu_{AB} +
\nu_{BC} + \nu_{CA} = 0$).  One calibration step of
\citet{baldwin1996} we do not need to apply is a photon-noise bias
correction to the triple product \citep[e.g.,][]{wirnitzer1985}, which
is only necessary for ``all-in-one'' combiners.

Pair-wise combiners (such as IONIC3) can have a large instrumental
offset for the closure phase which is generally calibrated by using
a point-source calibrator.  The instrumental closure phase of IONIC3
drifts less than 1~degree over many hours (a remarkable fact given the
unstabilized IOTA environment whose temperature can drift by $>$10~K
during a night), owing to the miniature dimensions of the IO
component.  Also, chromaticity effects limit our absolute precision
when the calibrator and target are not of the same spectral type due
to different effective wavelengths when using the broadband H filter.
Recent engineering tests indicate that the instrumental closure phase
($\Phi_{CP}$) of the current IONIC3 combiner varies little
($\simle$0.5\arcdegg) between between a hot star (B8) and a cool star
(M3) when using the broadband H filter. We minimize these errors
through repeated calibration of our targets with calibrators of
similar spectral type.  In light of this, we have adopted a systematic
error of $\Delta\Phi_{CP}=0.5$\arcdeg~ for the measurements presented
here to represent the residual correction for wavelength dependence.
This error is not applied in our analysis (it is not random and should
be fixed for each target/calibrator pair), but rather establishes a
minimum closure phase magnitude in order to be confident of detecting
{\em intrinsic} non-zero closure phase (i.e, skew) in a target. As a
cross-check we note that our targets are generally redder than our
calibrators, yet we have found no systematic offset in our measured
YSO closure phases, consistent with expectations from the engineering
study.  Lastly, we established the sign of our closure phase (baseline
triangle connecting telescopes
A$\rightarrow$B$\rightarrow$C$\rightarrow$A) using test measurements
of Capella and Matar in comparison to published orbits
\citep{hummel1994,hummel1998}.

\begin{figure}[hbt]
\begin{center}
\includegraphics[width=6in,clip]{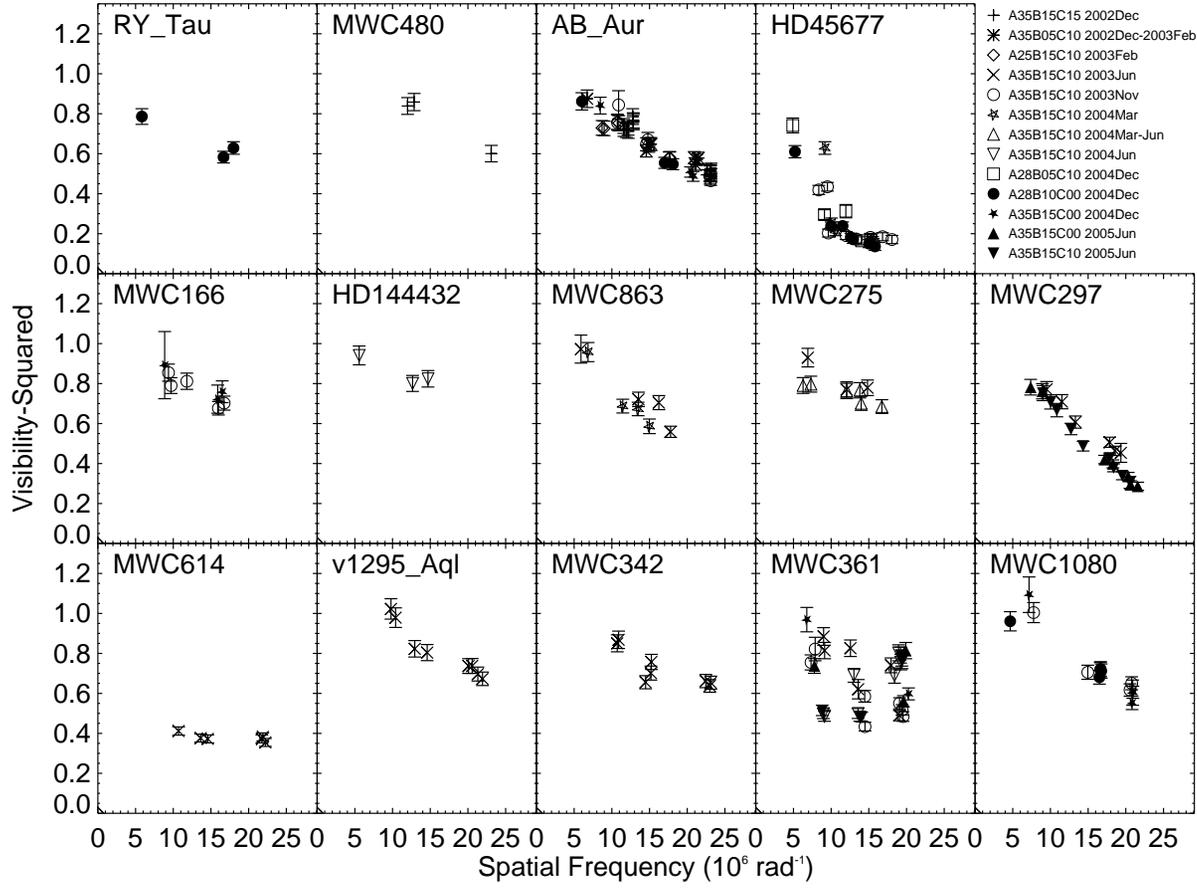}
\figcaption{\footnotesize These panels show the \viss results as a function of
spatial frequency for all 
targets presented in this paper.  Each observing configuration and
epoch are denoted by a unique plotting symbol according to the legend
in the upper-right panel.  
\label{fig_vis2}}
\end{center}
\end{figure}

In Figure~\ref{fig_vis2}, we present all the \viss data from this
survey work, where each epoch of observation is denoted by a unique
plot symbol. For this and subsequent data analysis, the \viss data
from a given configuration has been averaged in the uv-plane (on the
scale of uv-distance 4-meters) to improve signal-to-noise.  We have
compared our \viss results with those from previous workers and found
good agreement within quoted uncertainties.  Note that we report here
the first ever infrared interferometry results for the targets MWC~342
and HD~45677.

\begin{figure}[hbt]
\begin{center}
\includegraphics[width=6in,clip]{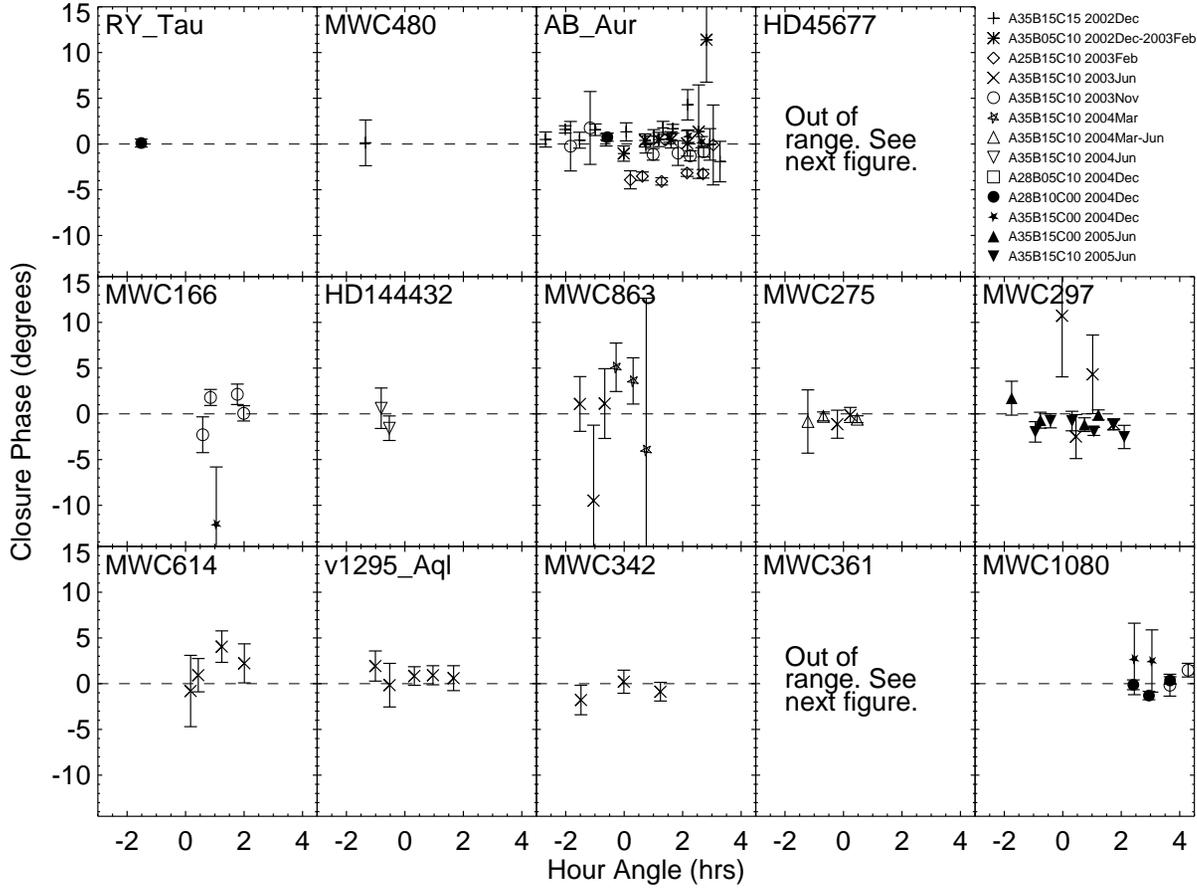}
\figcaption{\footnotesize These panels show the observed IOTA3 Closure Phase results
  as a function of hour angle for all targets presented in this paper.
  Each observing configuration and epoch are denoted by a unique
  plotting symbol according to the legend in the upper-right panel.
  Note that the closure phases for HD~45677 and MWC~361-A are too large to
  display here and are included in Figure~\ref{fig_cps2}
\label{fig_cps}}
\end{center}
\end{figure}

\begin{figure}[hbt]
\begin{center}
\includegraphics[angle=90,width=6in,clip]{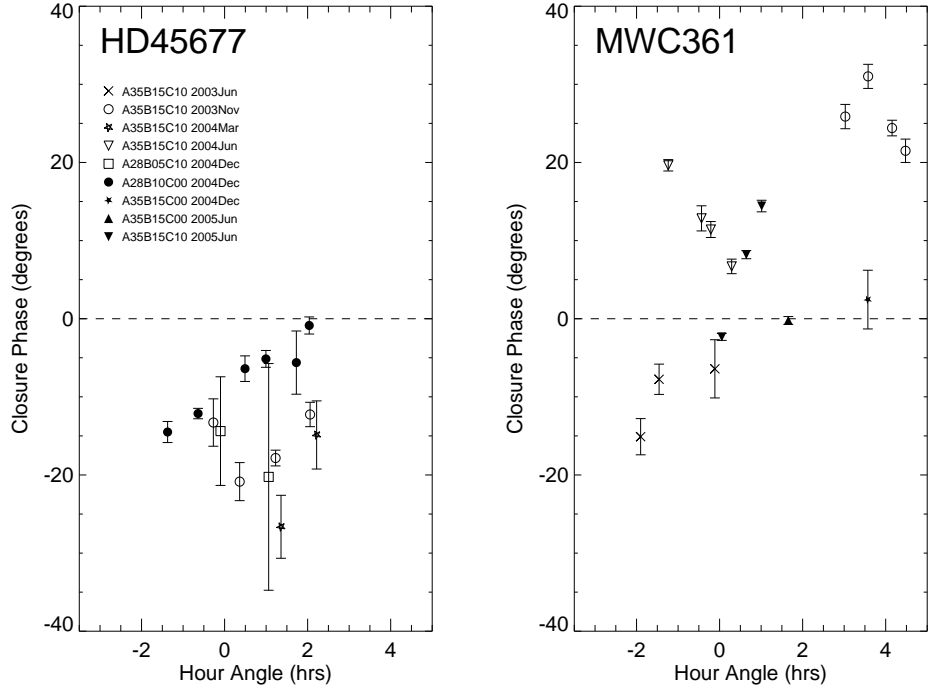}
\figcaption{These panels show the observed IOTA3 Closure Phase results
  as a function of hour angle for HD~45677 and MWC~361-A.
  Each observing configuration and epoch are denoted by a unique
  plotting symbol according to the legend in the left-hand panel.
\label{fig_cps2}}
\end{center}
\end{figure}

We present all the closure phase results in Figures~\ref{fig_cps} and
\ref{fig_cps2}, where the data has been plotted against the hour angle
of the observation and the array configuration is noted in the legend.
As was done for the \viss data, the closure phase data was 
averaged in the uv-(hyper)plane before plotting.

All \viss and closure phase data are available both in averaged and
un-averaged form from the authors\footnote{Currently, all data can be
  found at the data archives section of the Optical Long Baseline
  Interferometry News (OLBIN) website, http://www.olbin.jpl.nasa.gov};
all data products are stored in the FITS-based, optical interferometry
data exchange format (OI-FITS), recently described in
\citet{pauls2005}.

\section{Analysis of Visibility Data}
\label{section:results1}

\subsection{SED Decomposition and Size Estimation}
In this section, we carry-out the (now) standard procedure of YSO size
estimation, first outlined by \citet{rmg2001}.  While most of targets
have already been observed by either IOTA, PTI, or Keck, we have much
better (u,v)-coverage than previous studies as well as a few new
targets (MWC~342, HD~45677), necessitating a fresh visibility and disk
size analysis.

\begin{figure}[hbt]
\begin{center}
\includegraphics[angle=90,clip,width=6in]{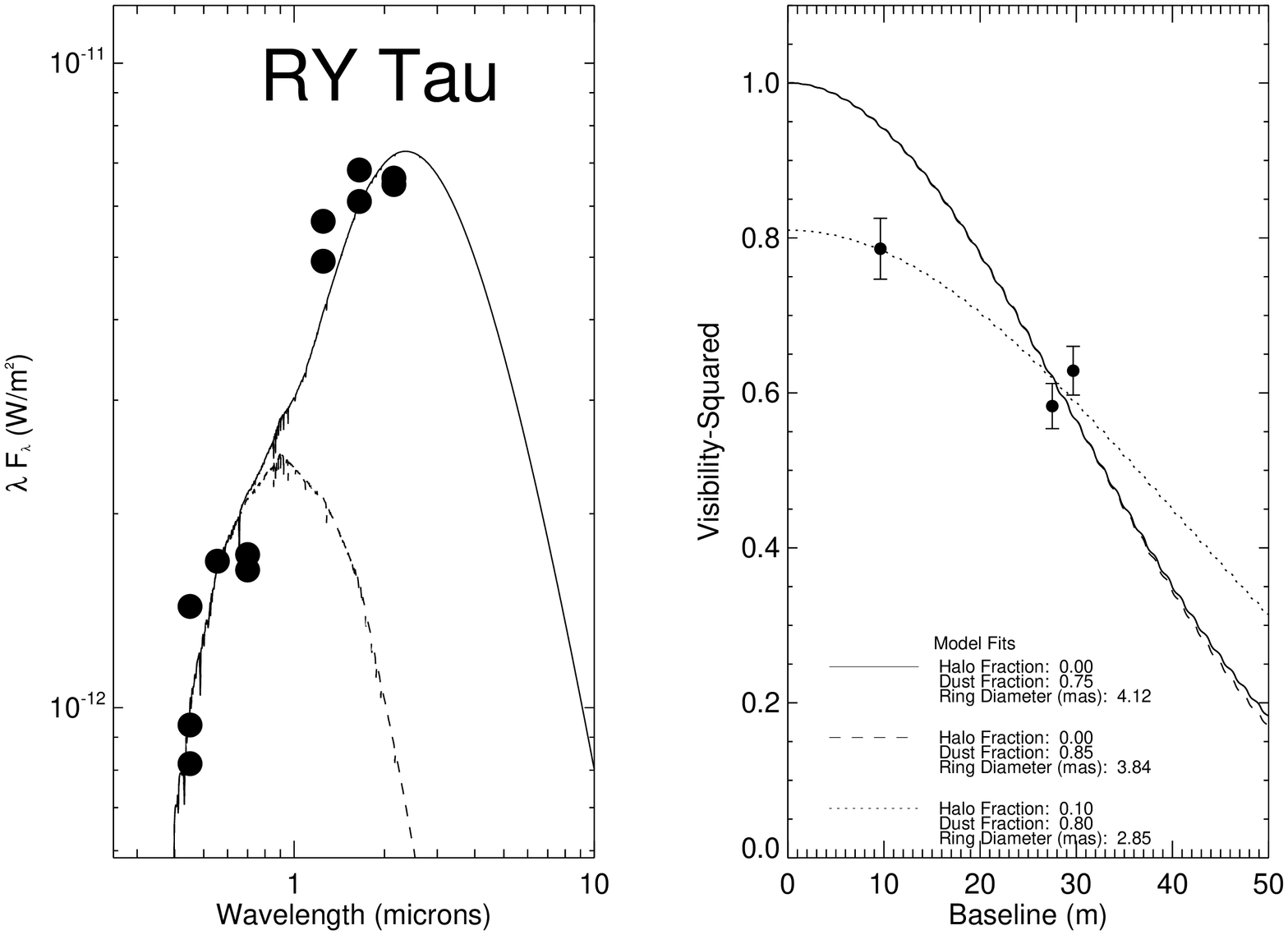}
\figcaption{(a) left-panel. Spectral Energy Distribution for RY~Tau
  including a simple two-component fit.  
 The dashed line represents
  the stellar contribution only and the solid line includes blackbody
  dust emission.  Photometry references included in
  Table~\ref{targets}.  No error bars are shown because intrinsic
  variability dominates over measurements uncertainty at any given
  epoch.  (b) right-panel. Visibility data are presented along with
best-fit ring models exploring different values for halo and 
dust fractions (see inset legend for details).
\label{sedvisRYTau}}
\end{center}
\end{figure}

\begin{figure}[hbt]
\begin{center}
\includegraphics[angle=90,clip,width=6in]{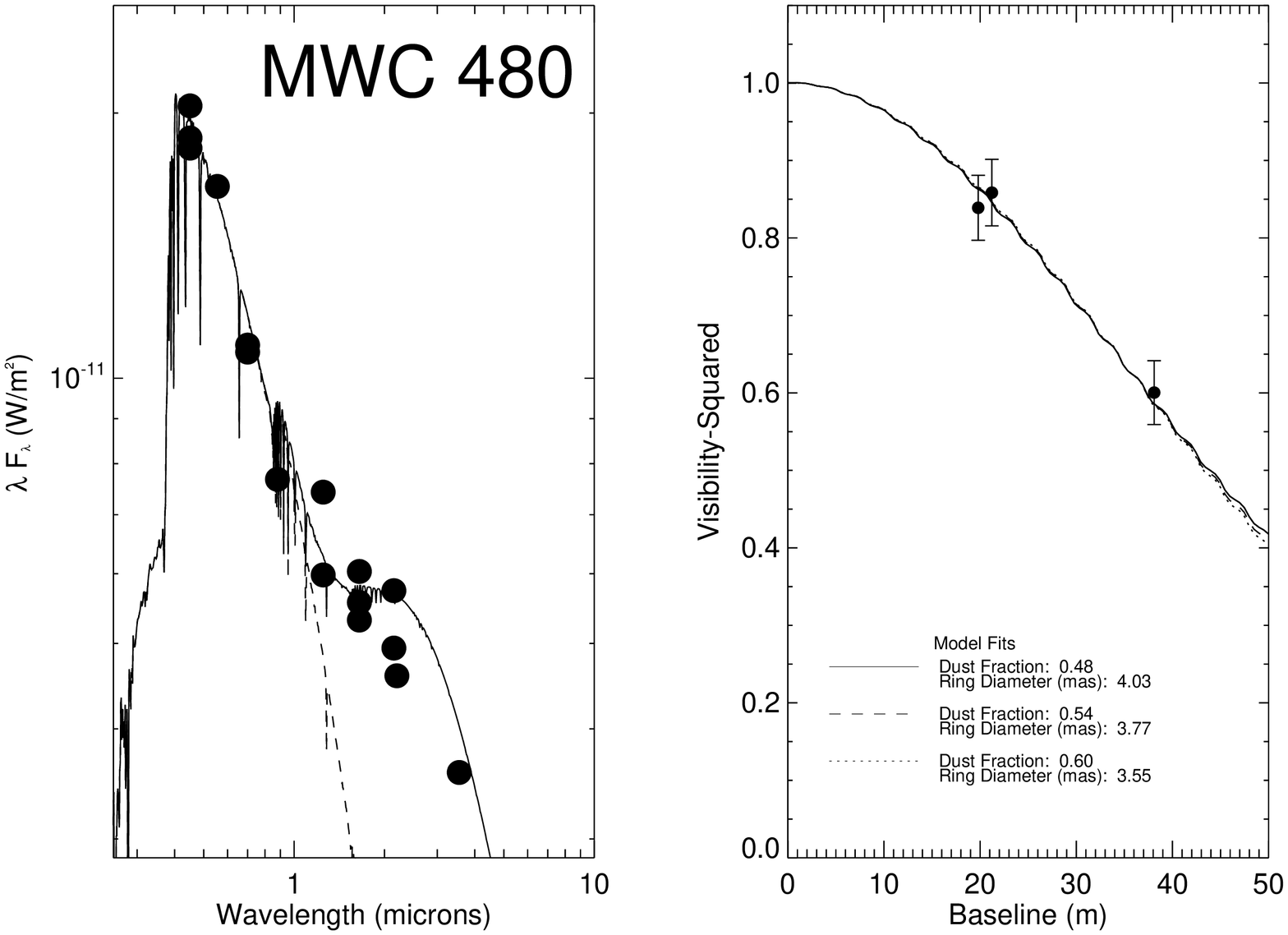}
\figcaption{(a) left-panel. Spectral Energy Distribution for MWC~480
  including a simple two-component fit.  The dashed line represents
  the stellar contribution only and the solid line includes blackbody
  dust emission.  Photometry references included in
  Table~\ref{targets}.  No error bars are shown because intrinsic
  variability dominates over measurements uncertainty at any given
  epoch.  (b) right-panel.  Multiple estimates for the dust fraction
  are used to constrain ring model fits to the IOTA3 visibility data.
\label{sedvisMWC480}}
\end{center}
\end{figure}

\begin{figure}[hbt]
\begin{center}
\includegraphics[angle=90,clip,width=6in]{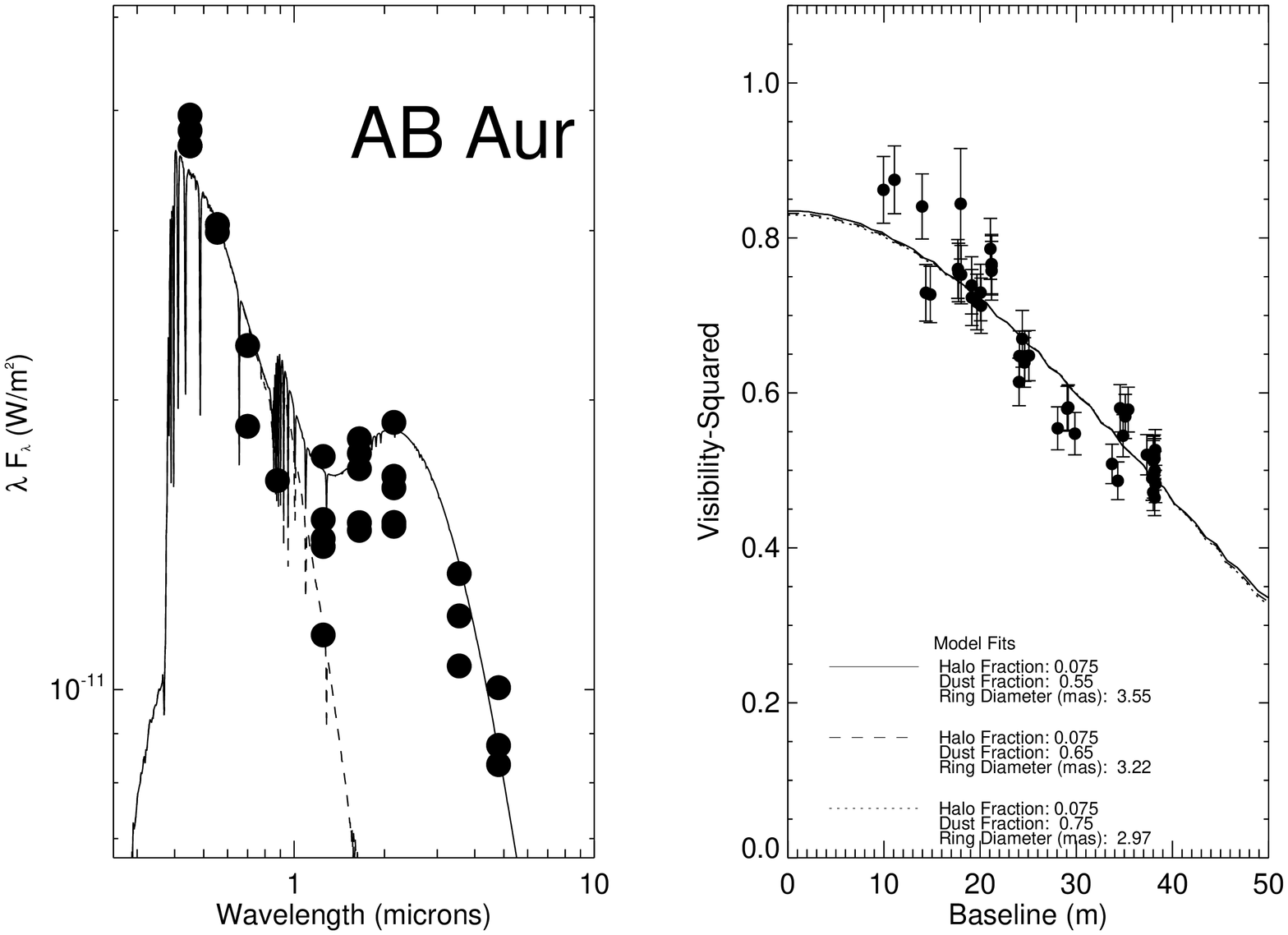}
\figcaption{(a) left-panel. Spectral Energy Distribution for AB~Aur including
a simple two-component fit.  The dashed line represents the stellar contribution
and the solid include blackbody dust emission.  Photometry references included
in Table~\ref{targets}.  No error bars are shown because intrinsic variability
dominates over measurements uncertainty at any given epoch.
(b) right-panel.  Multiple estimates for the dust fraction are used to constrain
ring model fits to the IOTA3 visibility data; here we also include 7.5\% emission
from an extended halo to fit short baseline data.
\label{sedvisABAur}}
\end{center}
\end{figure}

\begin{figure}[hbt]
\begin{center}
\includegraphics[angle=90,clip,width=6in]{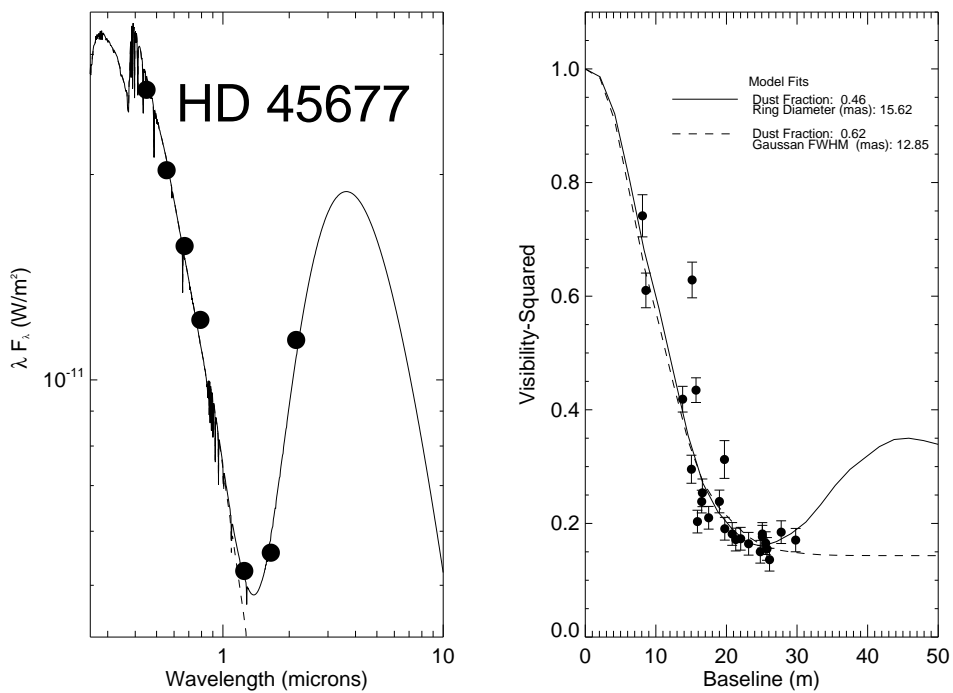}
\figcaption{(a) left-panel. Spectral Energy Distribution for HD~45677 including
a simple two-component fit.  The dashed line represents the stellar contribution
and the solid include blackbody dust emission.  Photometry references included
in Table~\ref{targets}.  Because this SED has changed significantly for this
target in recent years, we only include most recent photometry
(see Table~\ref{targets} for references).
(b) right-panel.  Here we fitted the IOTA3 visibility with a symmetric ring model
as well as a Gaussian emission model.  Large residuals here are a sign of elongated
asymmetric emission and a more complicated model is presented in Figure~\ref{modsHD45677}.
\label{sedvisHD45677}}
\end{center}
\end{figure}

\begin{figure}[hbt]
\begin{center}
\includegraphics[angle=90,clip,width=6in]{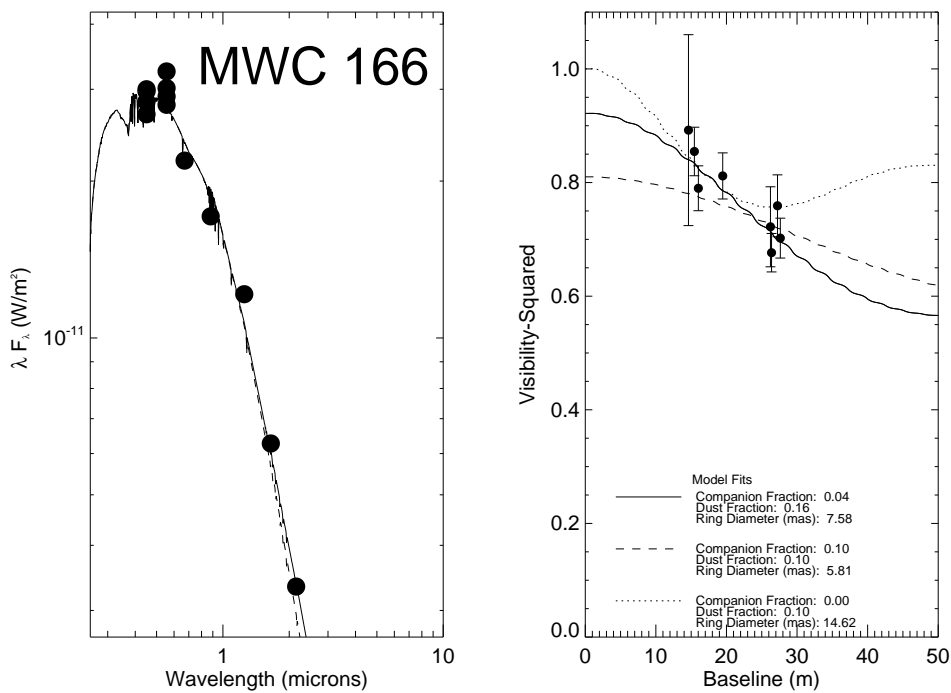}
\figcaption{(a) left-panel. Spectral Energy Distribution for MWC~166 including
a simple two-component fit.  The dashed line represents the stellar contribution
and the solid include blackbody dust emission.  Photometry references included
in Table~\ref{targets}.  No error bars are shown because intrinsic variability
dominates over measurements uncertainty at any given epoch.
(b) right-panel.  Because of a known, nearby companion which is completely
resolved in our observations, the visibility does not go to unity at the origin.
Multiple estimates for the companion contribution and dust fraction 
are used to constrain ring model fits to the IOTA3 visibility data.  
\label{sedvisMWC166}}
\end{center}
\end{figure}

\begin{figure}[hbt]
\begin{center}
\includegraphics[angle=90,clip,width=6in]{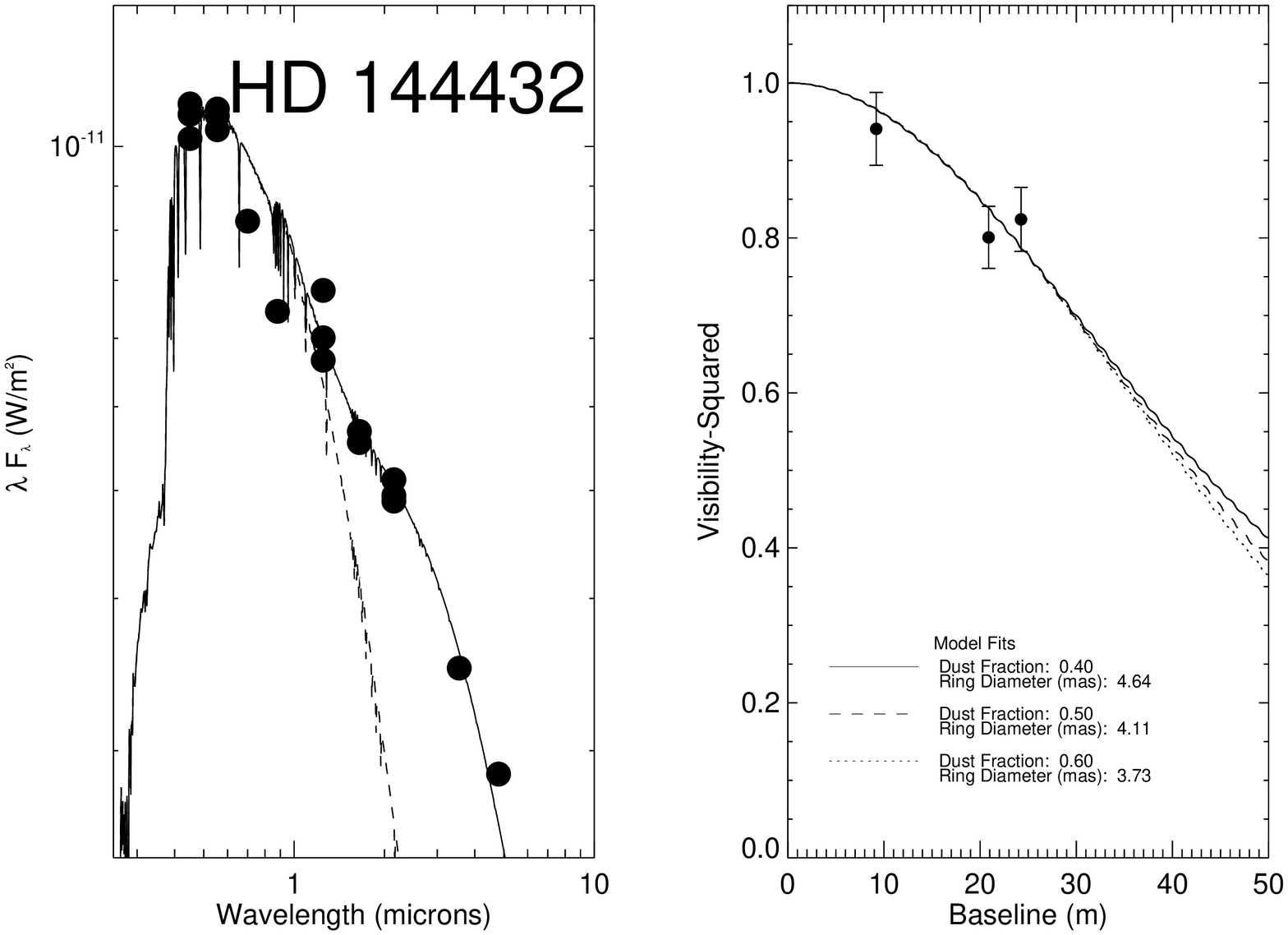}
\figcaption{(a) left-panel. Spectral Energy Distribution for HD~144432 including
a simple two-component fit.  The dashed line represents the stellar contribution
and the solid include blackbody dust emission.  Photometry references included
in Table~\ref{targets}.  No error bars are shown because intrinsic variability
dominates over measurements uncertainty at any given epoch.
(b) right-panel.  Multiple estimates for the dust fraction are used to constrain
ring model fits to the IOTA3 visibility data.  
\label{sedvisHD144432}}
\end{center}
\end{figure}

\begin{figure}[hbt]
\begin{center}
\includegraphics[angle=90,clip,width=6in]{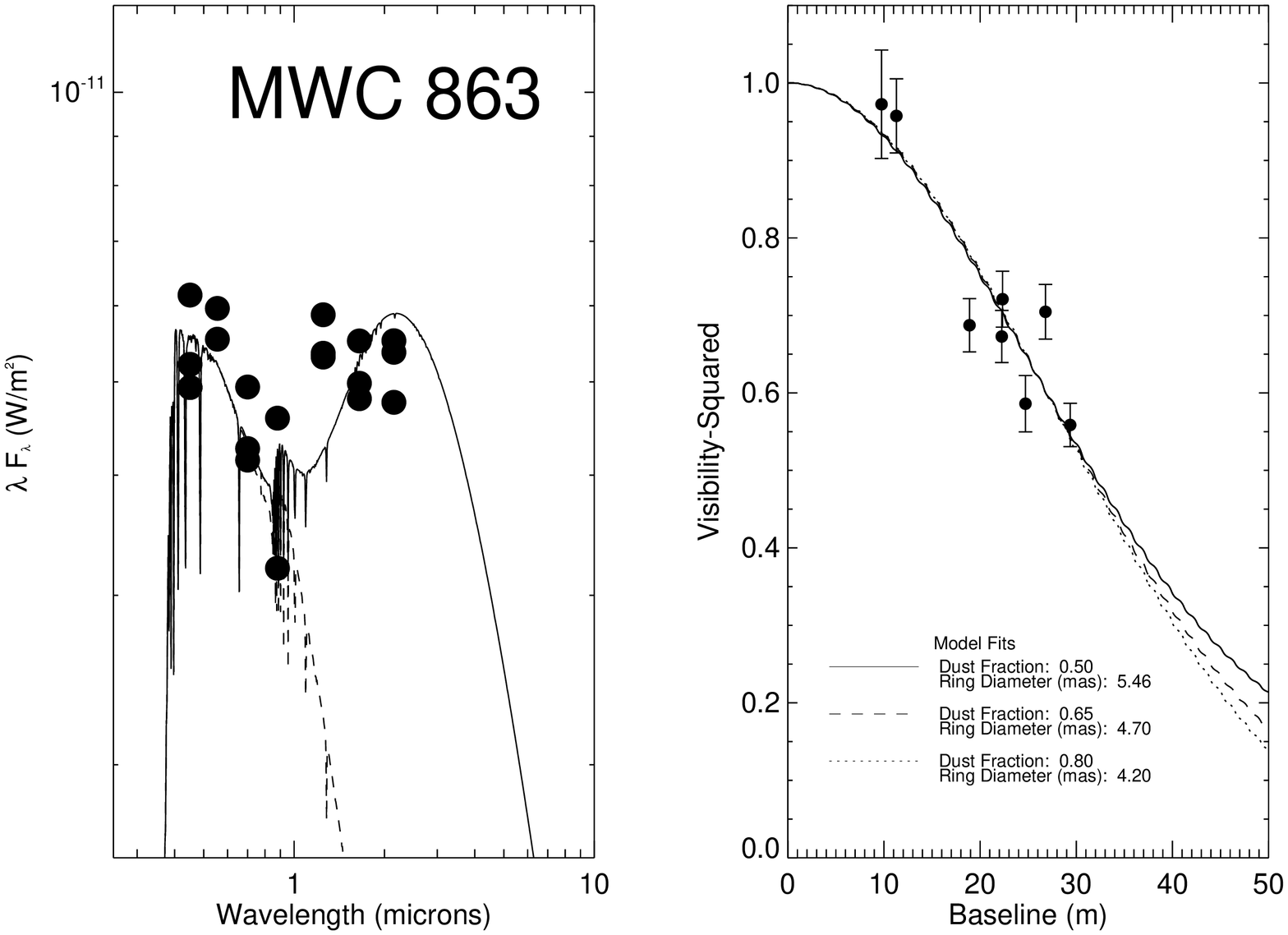}
\figcaption{(a) left-panel. Spectral Energy Distribution for MWC~863
  including a simple two-component fit, after removing
flux from the T~Tauri companion \citep{corporon1998}.  
 The dashed line represents
  the stellar contribution only and the solid line includes blackbody
  dust emission.  Photometry references included in
  Table~\ref{targets}.  No error bars are shown because intrinsic
  variability dominates over measurements uncertainty at any given
  epoch.  (b) right-panel. Visibility data are presented along with
best-fit ring models covering a large uncertainty range in
fraction of H band emission arising from disk.
\label{sedvisMWC863}}
\end{center}
\end{figure}

\begin{figure}[hbt]
\begin{center}
\includegraphics[angle=90,clip,width=6in]{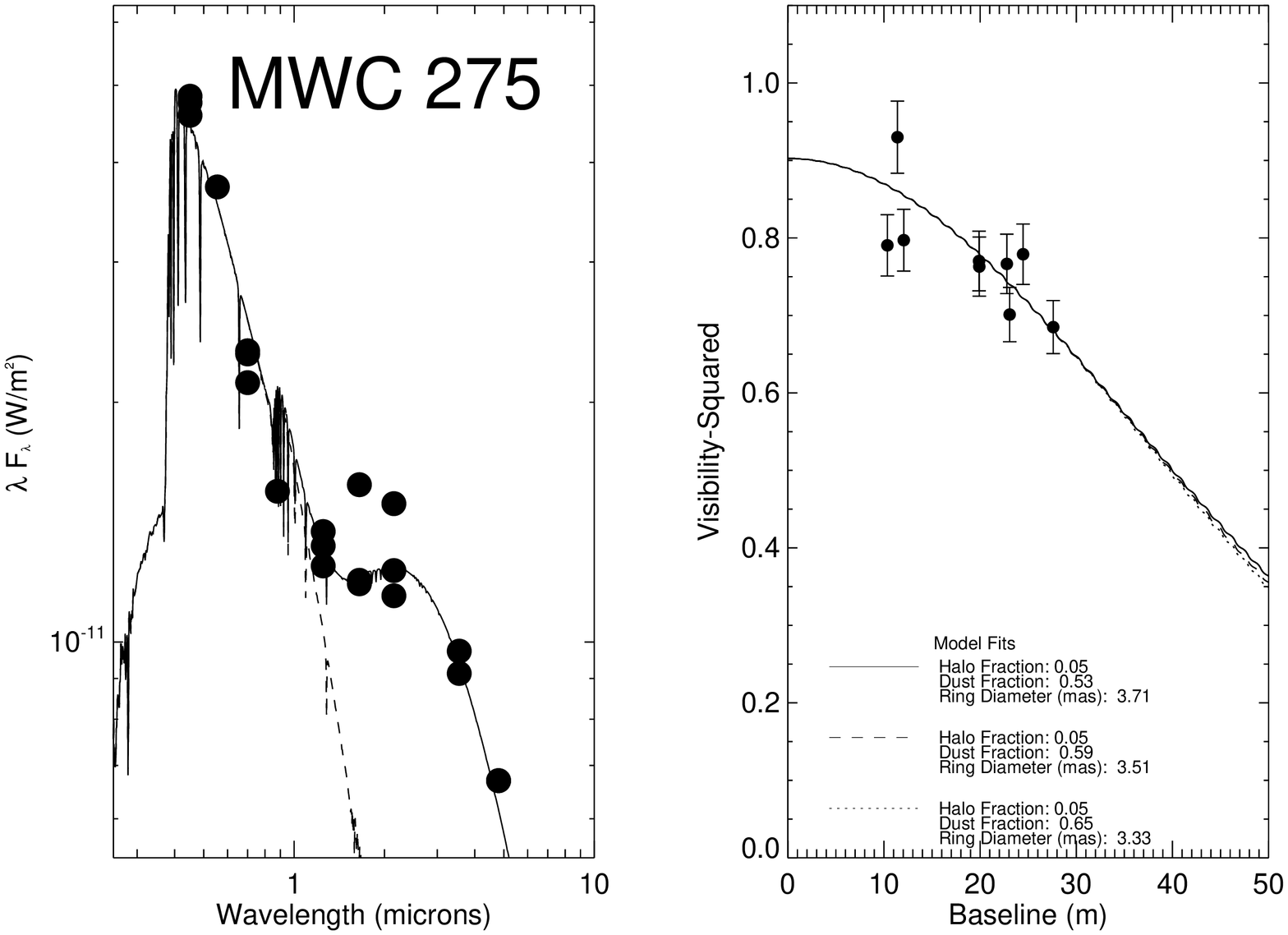}
\figcaption{(a) left-panel. Spectral Energy Distribution for MWC275 including
a simple two-component fit.  The dashed line represents the stellar contribution
and the solid include blackbody dust emission.  Photometry references included
in Table~\ref{targets}.  No error bars are shown because intrinsic variability
dominates over measurements uncertainty at any given epoch.
(b) right-panel.  Multiple estimates for the dust fraction are used to constrain
ring model fits to the IOTA3 visibility data; here we include 5\% emission from
extended halo to fit shortest baselines.  
\label{sedvisMWC275}}
\end{center}
\end{figure}

\begin{figure}[hbt]
\begin{center}
\includegraphics[angle=90,clip,width=6in]{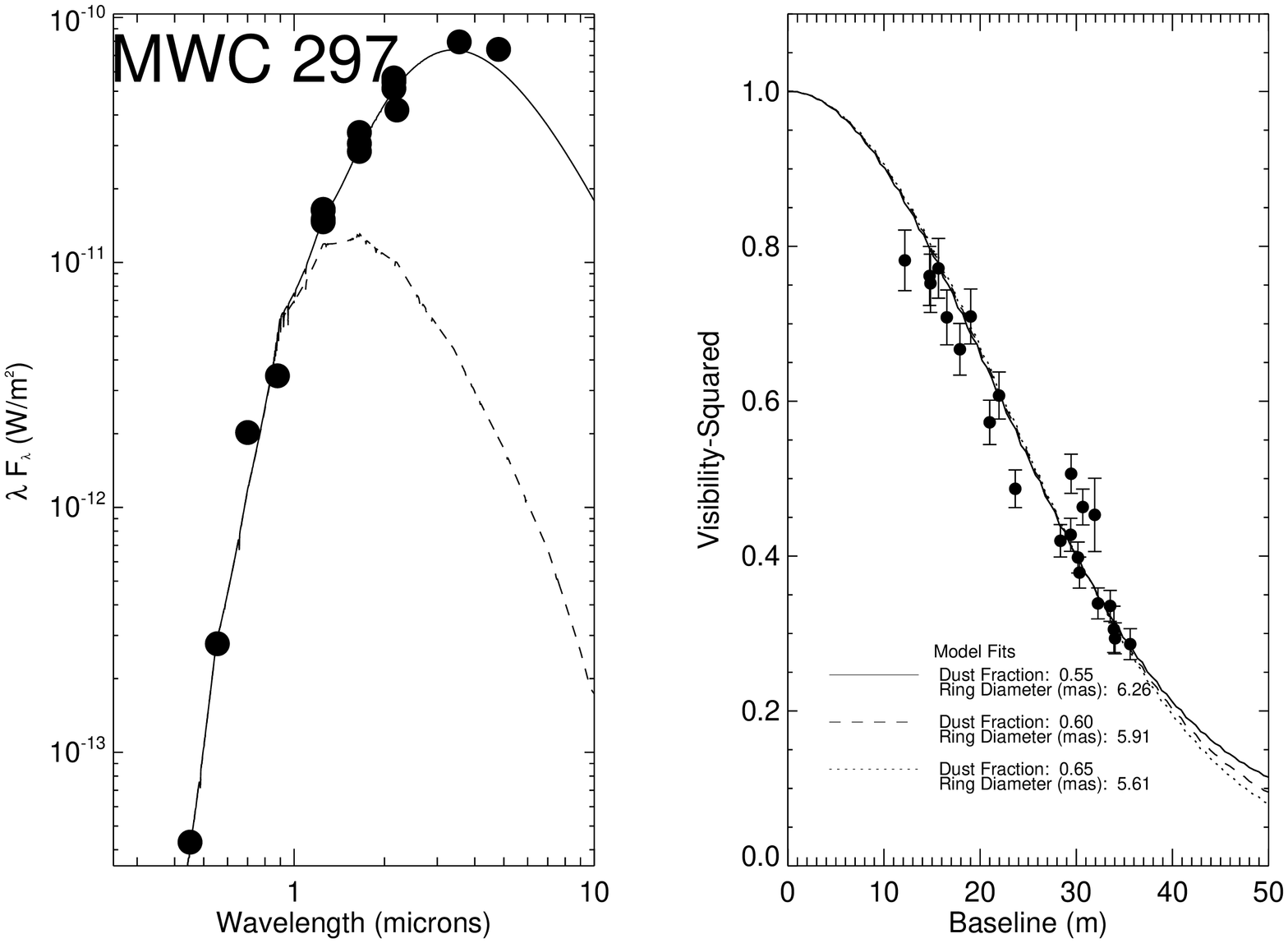}
\figcaption{(a) left-panel. Spectral Energy Distribution for MWC297 including
a simple two-component fit.  The dashed line represents the stellar contribution
only 
and the solid include both star and 
blackbody dust emission.  Photometry references included
in Table~\ref{targets}.  No error bars are shown because intrinsic variability
dominates over measurements uncertainty at any given epoch.
(b) right-panel.  Multiple estimates for the dust fraction are used to constrain
ring model fits to the IOTA3 visibility data.
\label{sedvisMWC297}}
\end{center}
\end{figure}

\begin{figure}[hbt]
\begin{center}
\includegraphics[angle=90,clip,width=6in]{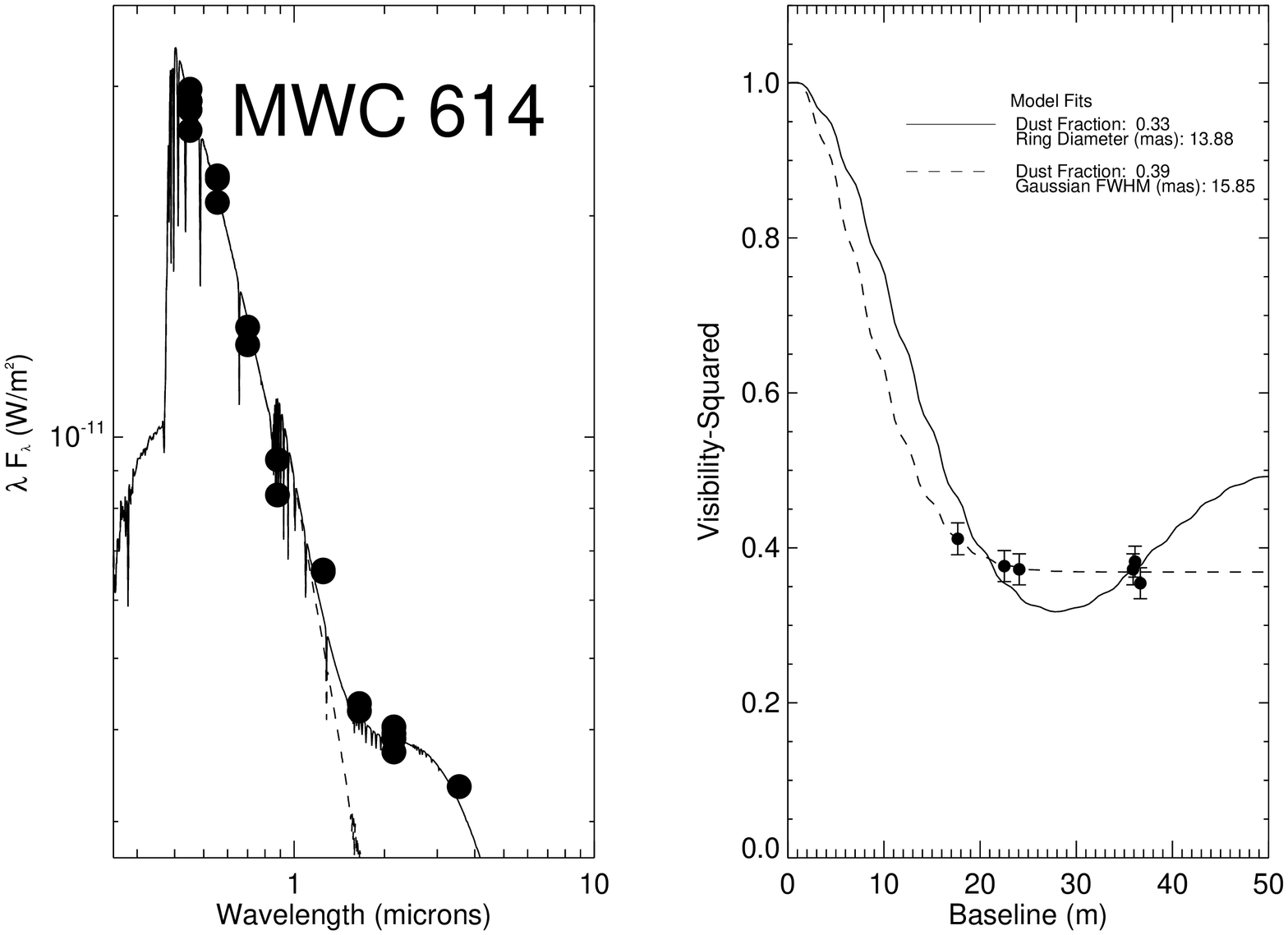}
\figcaption{(a) left-panel. Spectral Energy Distribution for MWC~614
  including a simple two-component fit.  The dashed line represents
  the stellar contribution only and the solid line includes blackbody
  dust emission.  Photometry references included in
  Table~\ref{targets}.  No error bars are shown because intrinsic
  variability dominates over measurements uncertainty at any given
  epoch.  (b) right-panel. Visibility data are presented along with
best-fit ring and Gaussian models.
\label{sedvisMWC614}}
\end{center}
\end{figure}

\begin{figure}[hbt]
\begin{center}
\includegraphics[angle=90,clip,width=6in]{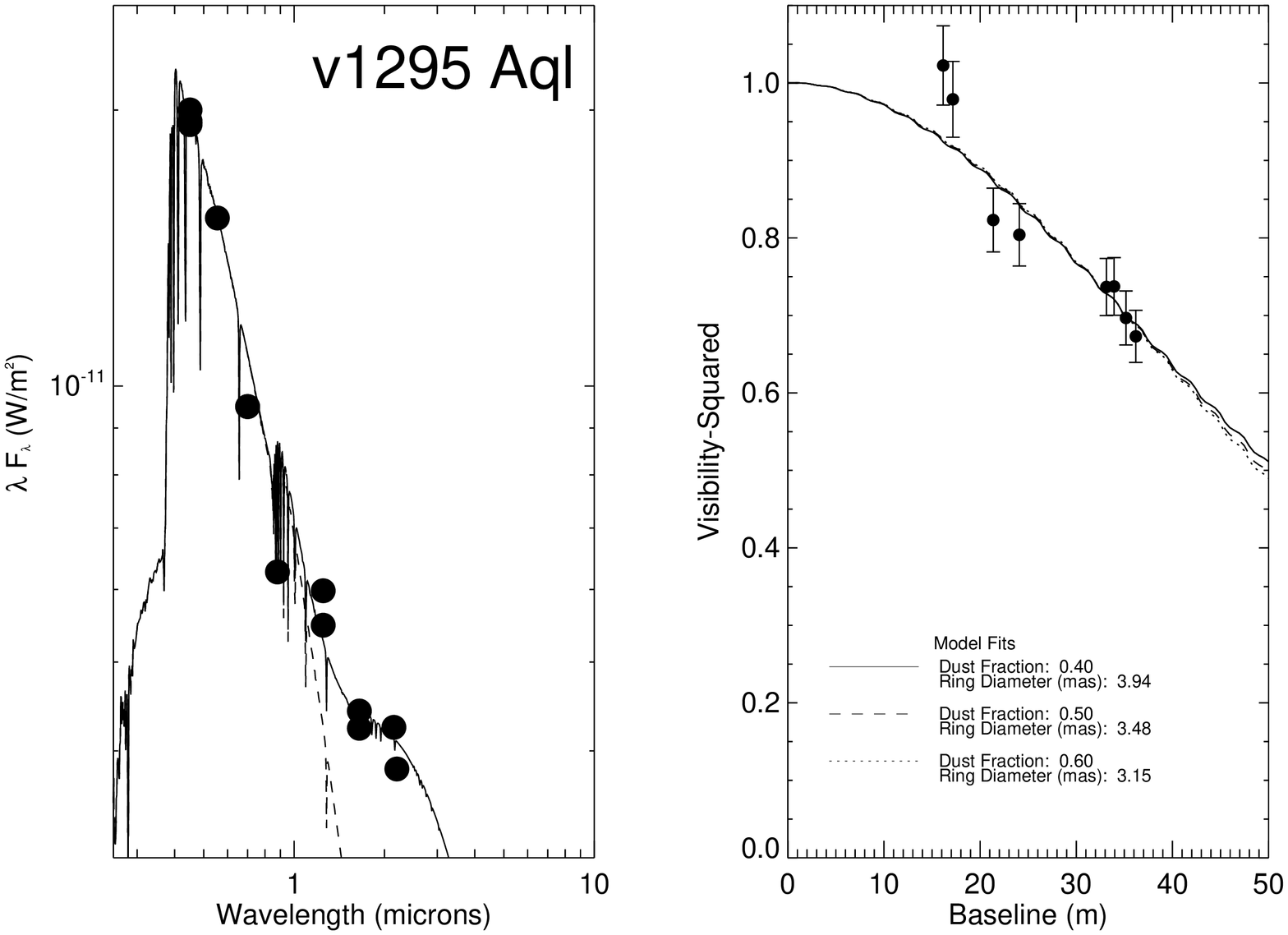}
\figcaption{(a) left-panel. Spectral Energy Distribution for v1295~Aql
  including a simple two-component fit, after removing
flux from the T~Tauri companion \citep{corporon1998}.  
 The dashed line represents
  the stellar contribution only and the solid line includes blackbody
  dust emission.  Photometry references included in
  Table~\ref{targets}.  No error bars are shown because intrinsic
  variability dominates over measurements uncertainty at any given
  epoch.  (b) right-panel. Visibility data are presented along with
best-fit ring models covering a large uncertainty range in
fraction of H band emission arising from disk.
\label{sedvisv1295Aql}}
\end{center}
\end{figure}

\begin{figure}[hbt]
\begin{center}
\includegraphics[angle=90,clip,width=6in]{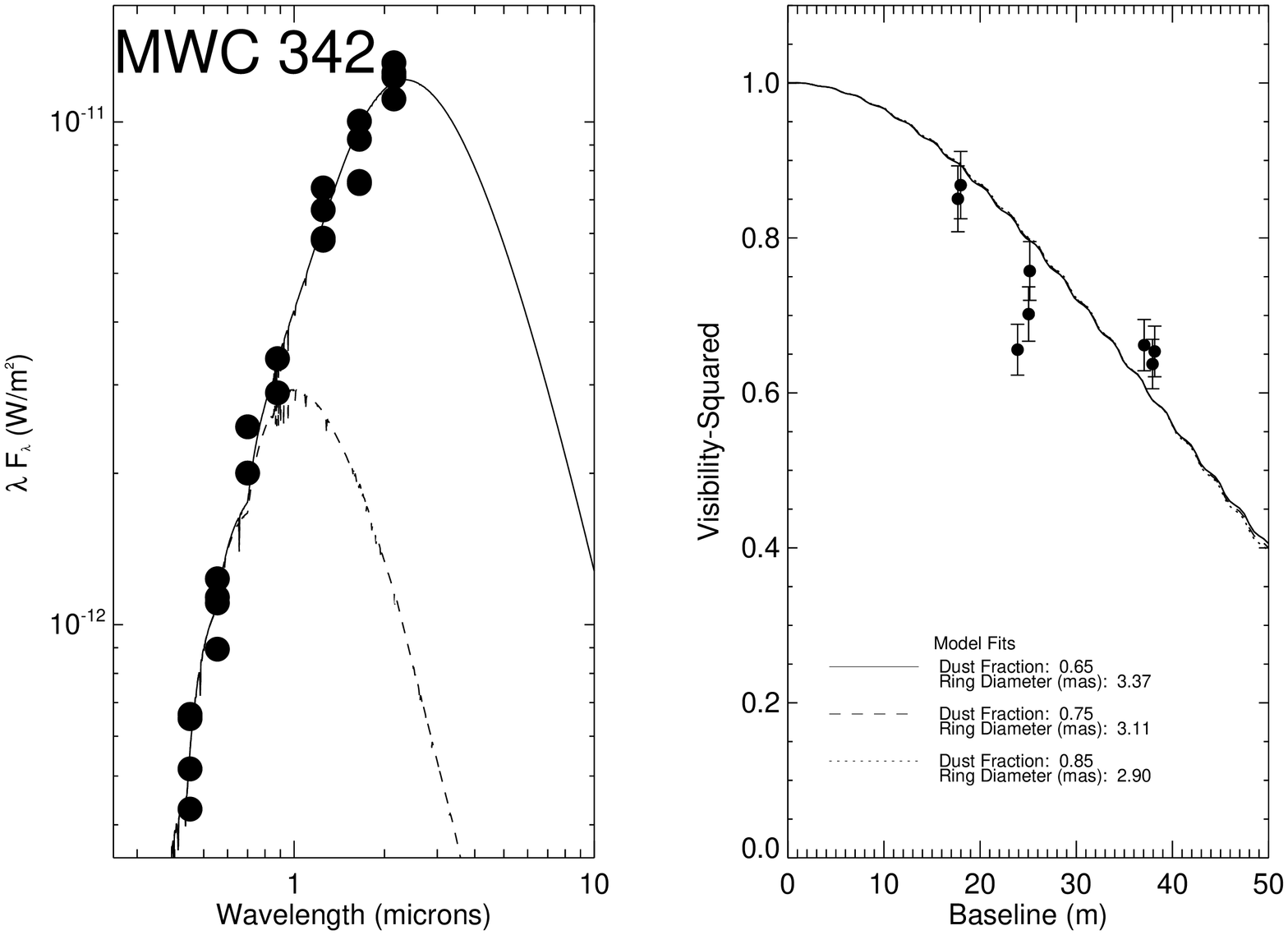}
\figcaption{(a) left-panel. Spectral Energy Distribution for MWC~342
  including a simple two-component fit.  The dashed line represents
  the stellar contribution only and the solid line includes blackbody
  dust emission.  Photometry references included in
  Table~\ref{targets}.  No error bars are shown because intrinsic
  variability dominates over measurements uncertainty at any given
  epoch.  (b) right-panel.  Multiple estimates for the dust fraction
  are used to constrain ring model fits to the IOTA3 visibility data.
 The poor fit might indicate elongated emission, however our limited
dataset does not allow further investigation  at this time.
\label{sedvisMWC342}}
\end{center}
\end{figure}

\begin{figure}[hbt]
\begin{center}
\includegraphics[angle=90,clip,width=6in]{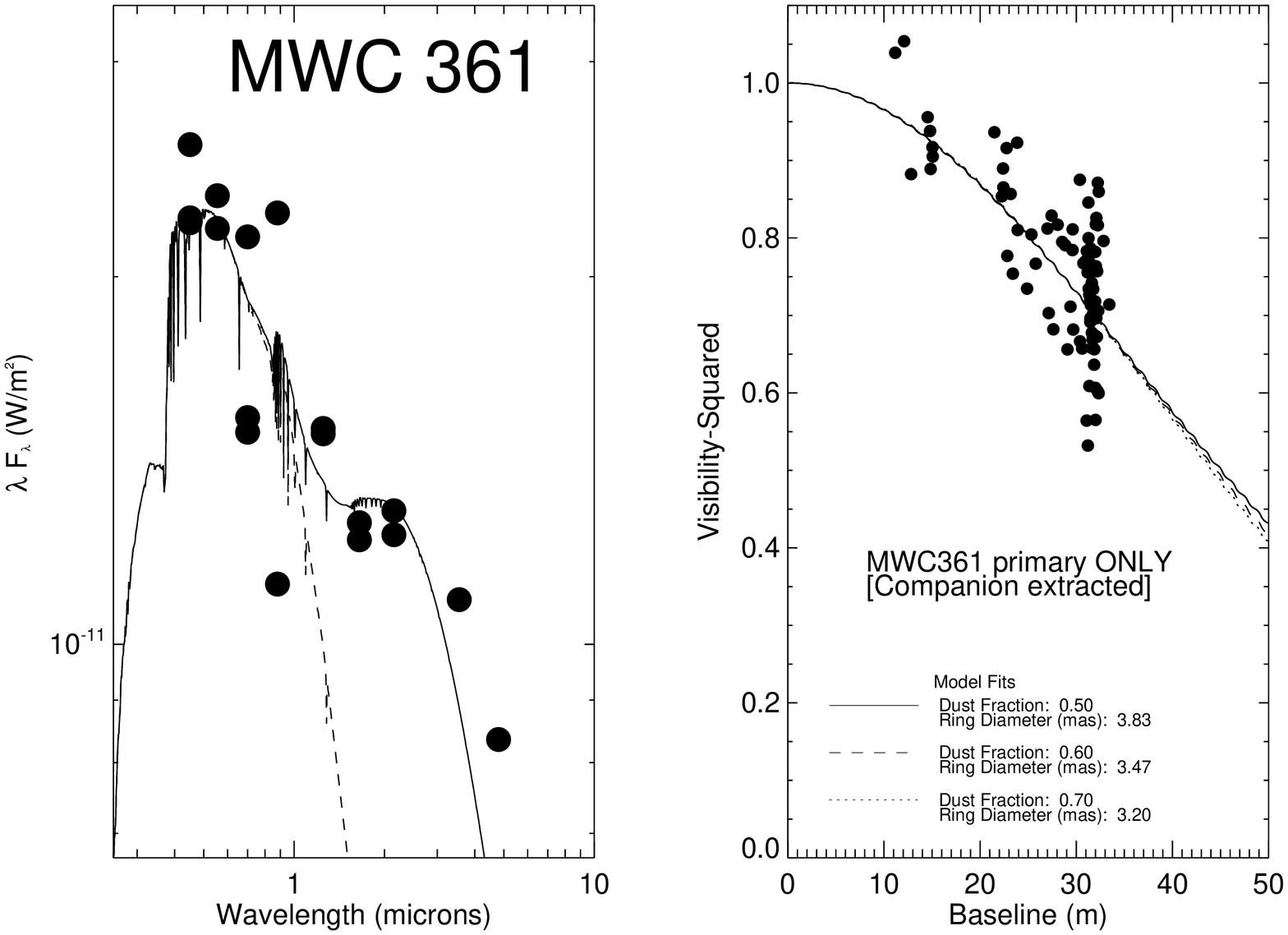}
\figcaption{(a) left-panel. Spectral Energy Distribution for MWC~361-A
  including a simple two-component fit.
 The dashed line represents
  the stellar contribution only and the solid line includes blackbody
  dust emission.  Photometry references included in
  Table~\ref{targets}.  No error bars are shown because intrinsic
  variability dominates over measurements uncertainty at any given
  epoch.  (b) right-panel.  Here, we present the corrected visibility data
of the primary component after removing the close binary companion.
No formal error bars are shown since the extraction process introduces an uncertain 
error, however the scatter in values should represent the uncertainty.
Visibility fits are presented along with
best-fit ring models covering a large uncertainty range in
fraction of H band emission arising from disk (based on SED decomposition).
\label{sedvisMWC361}}
\end{center}
\end{figure}

\begin{figure}[hbt]
\begin{center}
\includegraphics[angle=90,clip,width=6in]{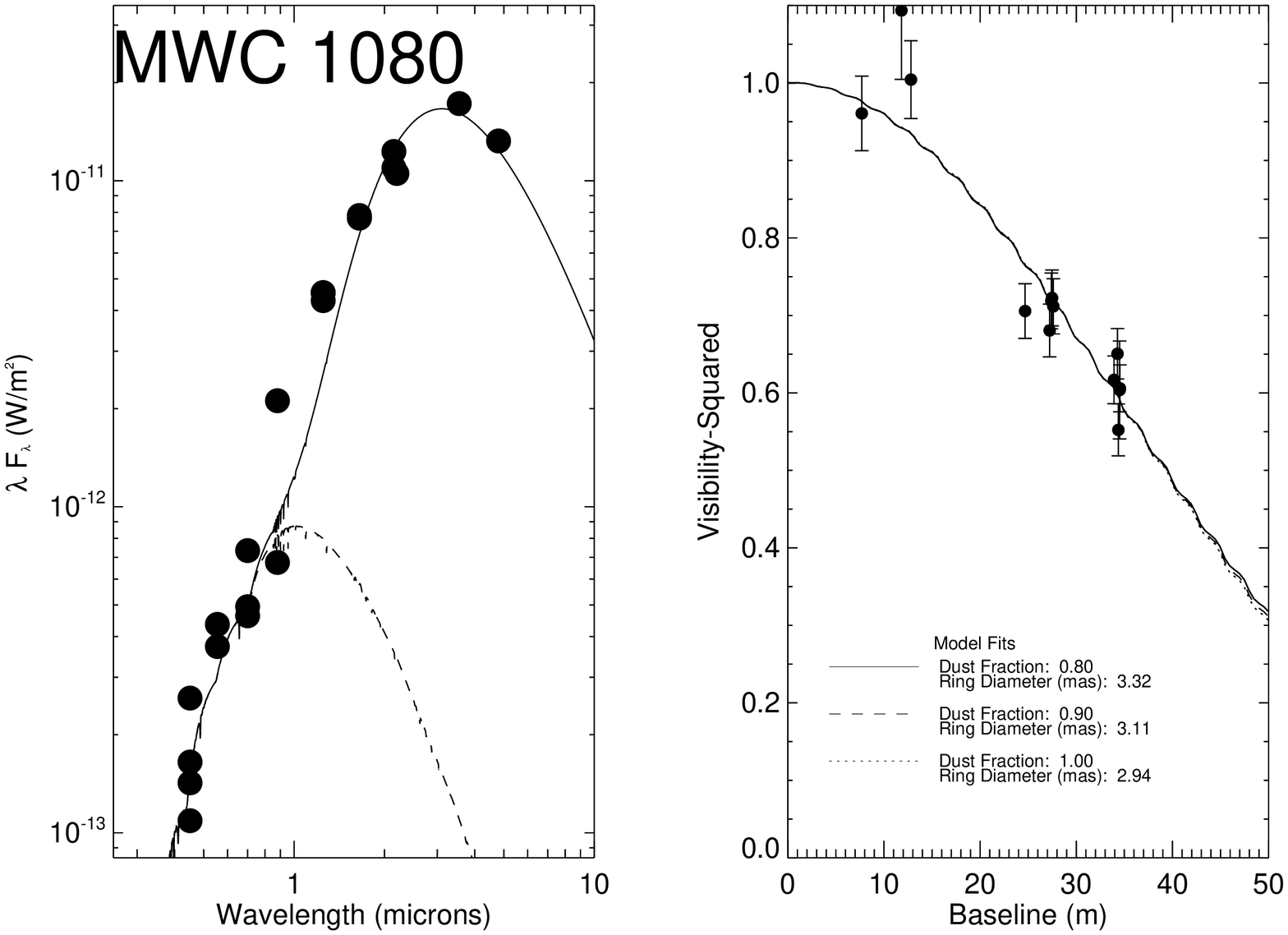}
\figcaption{(a) left-panel. Spectral Energy Distribution for MWC~1080
  including a simple two-component fit.  
 The dashed line represents
  the stellar contribution only and the solid line includes blackbody
  dust emission.  Photometry references included in
  Table~\ref{targets}.  No error bars are shown because intrinsic
  variability dominates over measurements uncertainty at any given
  epoch.  (b) right-panel. Visibility data are presented along with
best-fit ring models covering a large uncertainty range in
fraction of H band emission arising from disk.
\label{sedvisMWC1080}}
\end{center}
\end{figure}

Visibilities from YSOs are traditionally interpreted by fitting simple
emission models, such as Gaussians and rings
\citep[e.g.,][]{rmg2001,mm2002}, after correcting for the unresolved
stellar emission.  For each target, we have collected recent visible
and near-infrared photometry from the literature to create a spectral
energy distribution SED (see Table~\ref{targets} for list of specific
photometry references for each object).  Using the appropriate Kurucz
model \citep{kurucz1979} for each star (from literature spectral
type), we fit the SED as a combination of the star plus dust, the
latter being simply modelled as a single-temperature blackbody; the
reddening is another free parameter of this model important to match
the blue/visible portion of the spectrum.  After fitting, we can
estimate the fraction of flux at 1.65$\mu$m arising from the dust
component compared to the star.  Because YSOs are photometrically variable, we
included photometry from multiple epochs to illustrate the level of
variability and this uncertainty dominates the decomposition error
budget.   Figures~\ref{sedvisRYTau} to \ref{sedvisMWC1080} show the
results of the SED decompositions and our estimates of the dust
fraction (with errors) are tabulated in Table~\ref{diskresults2}.  A
more detailed description of this algorithm can be found in
\citet{monnier2005}.

Next, the \viss data are fit using a two-component model consisting of
the unresolved stellar source (flux fraction constrained from SED
fitting) and a thin circularly-symmetric ring of NIR emission
(Gaussian cross-section, 25\% thickness)\footnote{While our new observations
  have better (u,v)-coverage than previous work, the relatively short
  baselines of IOTA coupled with the 5\% calibration errors on the
  visibility-squared forbid us from definitively detecting large-scale
  elongations (for most sources) such has been reported using the
  longer-baseline Palomar Testbed Interferometer
  \citep[e.g.,][]{eisner2004}.}.  In some cases, a fully-resolved
extended component (``halo'') was included if evident from the
data. These results are found in Table~\ref{diskresults2} and the
right panels of Figures~\ref{sedvisRYTau}-\ref{sedvisMWC1080}.
Note that this ``halo'' need not be of the kind 
described by \citet{vinkovic2006}, but could
arise form scattering off the surface of the disk, or from 
localized emission within even a few AU of the star \citep{rmg2006}.

In order to interpret the presence or absence of ``halos,'' one must
understand the field-of-view of modern infrared interferometers.  The
most important limitation on field-of-view (FOV) is restricted by the
field of an individual telescope.  For fiber-optic based systems
(Keck, PTI, IOTA3-IONIC), this FOV corresponds to the
diffraction-limited beam which is 1.0$\arcsec$ for IOTA3-IONIC at
H-band, 1.4$\arcsec$ for PTI at K-band, and 0.055$\arcsec$ for Keck at
K-band. Note the original IOTA2 experiment \citep{rmg2001} had a
``free-space'' combiner which had a 2-3$\arcsec$ FOV.  Thus, light
beyond this scale does not make it into the beam combiner and is
completely neglected.  The {\em interferometric} FOV is usually
limited by bandwidth-smearing.  This means that light farther than
$\frac{\lambda^2}{\Delta\lambda~B}$ ($B$ is the interferometer
baseline, $\lambda$ is the central observing wavelength, and
$\Delta\lambda$ is the bandwidth) from the central source becomes
incoherent and does not produce fringes at the phase center (zero
optical path length difference).  In practice, this is about 6$\times$
the fringe spacing, corresponding to about 60~mas for IOTA (H-band),
25-30~mas for PTI \& Keck (K-band). In this latter case, light outside
the interferometric FOV is detected but is not coherent and acts to
lower the observed visibility. See \citet{monnier2003} for further discussion of
modern optical and infrared interferometry.

 Before discussing individual objects, we note that evidence for
 large-scale (0.01-0.50$\arcsec$) ``halos'' was found for 
 AB~Aur \citep[as found by][]{rmg2006}, MWC~275, MWC~297, and
 RY~Tau. Also, we surprisingly found that the NIR disk emissions for
 HD~45677 and MWC~614 were {\em completely resolved} by the 38~m IOTA
 baselines; this allowed a unique comparison of the {\em predicted}
 stellar flux contribution (from SED fitting) with the observed values
 as well as search for the tell-tale signature of {\em ring-like}
 emission compared to more smooth {\em Gaussian-like} emission.  We
 discuss unique aspects of some targets below.

\subsection{Individual objects}
\label{individual}

Many objects need special attention and specific discussion.  In this section we
briefly describe unusual and interesting characteristics of targets in our sample.

{\bf RY~Tau.} This star is variously spectral-typed as K1
\citep{akeson2005} to F8 \citep{mora2001}, an intermediate object
between the Herbig Ae/Be and Classical T~Tauri stars.  Adopting F8 and
fitting to the SED, we find a stellar luminosity of $\sim$5~\lsun, a
few times lower than reported by other workers.  The quality of
the RY~Tau visibility data is particularly excellent owing to good seeing, and we
take seriously the low visibility data point at short baselines seen in
Figure~$\ref{sedvisRYTau}$.  In this figure, we present fits both
assuming a 10\% extended ``halo'' emission and with no halo emission (both
results are also reported in Table~\ref{diskresults2}).  Clearly, halo
emission fits data much better and gives a ring diameter consistent
with K band results of \citet[][ring diameter
2.60$\pm$0.14~mas]{akeson2005}, although this high level of H-band extended emission seems
incompatible with their reported 2\% scattered light at 0.9$\mu$m.

{\bf AB~Aur.}  This object was discussed in detail in \citet{rmg2006}. Here,
we present an independent size analysis, but use the same IOTA dataset.

{\bf HD~45677.}  Here we report the first long-baseline interferometry
data for this target.  To aid interpretation, we obtained 
recent visible photometry from
the MDM Observatory in 2004 November and used NIR photometry from 2MASS. 
Surprisingly, the disk emission appears fully resolved at the longest IOTA
baselines, a direct measurement of 62$\pm$3\% dust fraction, in 
reasonable agreement with the value derived from the 
SED decomposition (65$\pm$15\%). 

Interestingly, the visibility data does not show the expected shape
for ring-like emission (specifically, large ``ringing'' at long
baselines due to the ring thinness).  Although we were able to find an
acceptable fit for a ring model, the apparent constant visibilities at
long baselines motivated additional fitting of a Gaussian emission
model and this result, along with the best-fit ring model, is included
in Table~\ref{diskresults2} and Figure~\ref{sedvisHD45677}.  As is
obvious from this figure, additional data at slightly longer baselines
can rule out or definitively establish ring-like emission (as opposed
to smooth, Gaussian emission) for this object.  The distinction may be
critical to decide between envelope and inner-rim models of the NIR
emission \citep[e.g.,][]{vinkovic2006}.

Here, we first note that the emission appears quite elongated and we
additionally find large non-zero closure phases. This target is
clearly very different from the others in our sample, and a more
complex model is developed in \S\ref{hd45677cp} after a more general
discussion of closure phases.  

{\bf MWC 166.} \citet{rmg2001} found that there is very little, if any, IR
excess at H band ($12\pm20$\%) for this high-luminosity target (B0).
However, these workers did find the long-baseline visibility to be
resolved  (V$\sim$90\%), after correcting for 22\% emission from
nearby companion \citep[also B0 spectral type at separation 
$\sim$0.6$\arcsec$,][]{corporon1998}.  
We have obtained new visible photometry at MDM
Observatory which, when combined with 2MASS data, also suggest a small IR
excess ($10\%\pm10\%$).  

Our new visibility data also confirm the slightly resolved visibility
at long-baselines (V$\sim$85\%), although this is {\em before}
correction for the known companion.  Because the IONIC3 combiner uses
single-mode fibers (unlike the original IOTA two-telescope survey),
the companion flux is more attenuated compared to the primary star,
although the exact amount is difficult to predict (depending on the
seeing and the optical alignment). At
short baselines (15~m), the observed visibility is $\sim$90\% -- this
limits the companion's contribution to $\leq$10\% of total H band
flux.  In order to account for this, we performed fits including both
0\% and 10\% contribution from companion, while also constraining the
dust fraction to be between 1-20\%.  The allowed ring diameters span
the range 4-15~mas, with a best-fit for 80\% primary star, 4\%
secondary star (incoherent flux), and 16\% disk emission using a ring
diameter of 7.5~mas.  Some representative fits are shown in
Figure~\ref{sedvisMWC166}.

{\bf MWC~863.} A binary companion was detected by
\citet{corporon1998}, a T~Tauri star (K4) at 1.1$\arcsec$ and $\Delta
H\sim$2.24. We see no evidence for this companion in our visibility
data, since the field-of-view of an IONIC3 single-mode fiber is
only $\pm$0.5$\arcsec$.  Although the companion flux was taken into
account in our estimate of the dust fraction in SED fitting, the SED
fit is still poor in the near-IR region.  While our measured \viss is
similar to that found by \citet{rmg2001}, our ring diameter estimate
is much smaller because we did not apply the same visibility
``correction'' to compensate for the incoherent flux of the companion
\citep[our result here is consistent with the recent result using the
longer baseline Keck Interferometer at K band;][]{monnier2005}.  Also,
our luminosity determination is significantly lower here than recently
reported in \citet{monnier2005}, due to selection of more recent
optical photometry.  Better coeval photometry is needed for this
target in order to estimate stellar and disk properties 
precisely.

{\bf MWC 297.} \viss data are better fit using a 5\% extended ``halo'' emission,
although the fit is reasonable assuming no halo (see
Figure~\ref{sedvisMWC297}).  The best-fit values (and uncertainty
estimates) in Table~\ref{diskresults2} include a 2.5$\pm$2.5 \% estimate
for halo emission. 
This is one of the most resolved disks in our survey and,
while our uv-coverage is limited, there is slight indication of elongation along
PA$\sim$100$\arcdeg$; more data are needed to confirm this.

{\bf MWC~614.} The nearly flat visibility curve indicates the dust
emission is entirely resolved even at short baselines, and that the
star makes up 61\% of the flux, consistent with SED decomposition
estimate of 63$\pm$10 \%. We can convert this to a lower limit on the
dust envelope size, ring diameter $\simge$13~mas (best-fit Gaussian
FWHM 15.9~mas; see Figure~\ref{sedvisMWC614}).  However, as for
HD~45677, we see no telltale sign in the \viss data for a {\em
  ring-like} structure which has the distinctive Bessel function shape
in Fourier-space (e.g, Gaussian profile is a  better fit than ring
profile).  This target would be ideal source for adaptive optics on an
8-m class telescope. The large ring size derived here is anomalous for
the Herbig Ae size-luminosity relation \citep{monnier2005}. This could
be partially-explained if there is a stellar companion within about
1$\arcsec$ of the primary star, since this would lead to an
overestimate of the dust ring size in our study.

{\bf MWC 342.}  We present first long-baseline interferometry data on this
target.  While our data indicates elongated emission, our 
very limited data set does not allow us to meaningfully quantify this. 
We note that the evolutionary state of MWC~349 is unclear; 
\citet{mirosh1999} suggests this star, as well as HD~45677, are possibly 
X-ray binaries like CI~Cam.

{\bf MWC~361-A.} A 6$\arcsec$ companion to MWC361-A was first discovered
by \citet{li1994}. While this object does not affect our measurements,
the close companion (separation $\sim$18~mas) discovered by
\citet{rmg2001} obviously is easily detected and appears to be related
to the $\sim$3.7~yr periodic variations in the H$\alpha$ emission seen
by \citet{mirosh1998}. Indeed, \citet{pogodin2004} recently reported a
detailed spectroscopic orbit (more discussion of this in
\S\ref{mwc361orbit}).  Here in Figure~\ref{sedvisMWC361}, we have
plotted the \viss of primary star after correcting for the fainter
companion contribution (made possible only after the full orbit
analysis of \S\ref{mwc361orbit}). Note that this ring diameter
estimate assumes the companion star is itself {\em unresolved}, a
reasonable assumption based on the observed size-luminosity relations.
Interestingly, the NIR ring diameter found here is undersized compared
to the normal size-luminosity relation for Herbig Ae/late Be stars, as
also found for some other early Be Herbigs \citep{mm2002,eisner2004}.

{\bf MWC~1080.}  There seems to be a fair amount of uncertainty on the
spectral type of this star, ranging from B0-2 \citep{cohen1979} to
A0-3 \citep{yoshida1992}, and distances of 1 to 2.5~kpc.  Here, we
adopt the earlier spectral type and 
a distance of 2.2~kpc \citep{grankin1992}.  In principle, our
visibility data should be sensitive to the ``wide'' binary
\citep{corporon1998} seen with a separation of $\sim$0.8$\arcsec$
($\Delta H\sim2.8$), however there is no sign of this in our data, 
presumably due to high brightness ratio and relatively large off-axis distance.

\section{Closure Phase as Tool for Studying YSOs}
\label{cpintro}
\subsection{Background}
Basic interferometry theory informs us that an interference fringe
measured on an individual baseline corresponds to a single Fourier
component of the brightness distribution under scrutiny.  Basic
Fourier theory further informs us that any brightness distribution can
be reconstructed by collection of sufficient Fourier components (i.e.,
``visibility'' data).  Thus, an ``imaging'' interferometer, such as
the radio Very Large Array (VLA), is characterized by a large number
of telescopes (corresponding to many simultaneous Fourier components)
and a range of baselines (for imaging a range of spatial scales).
While the VLA accomplishes this using 27~telescopes with a maximum
spacing/minimum spacing of $\simge$100, current infrared interferometers
have 3-6 telescopes with maximum/minimum spacings of $\simle$10.  Not
surprisingly, current interferometers have difficulty performing
model-independent image reconstructions (except for simple scenes such
as binary stars).

The problem is made more vexing for optical interferometry since
atmospheric turbulence scrambles the Fourier {\em phases} of the
observed interference fringes.  In some cases, these Fourier (fringe)
phases are not important -- it can be easily proven that {\em
  centrosymmetric} images can be completely characterized by the
fringe amplitudes alone\footnote{By defining the coordinate origin at
  the center of symmetry, all the Fourier phases become 0 or $\pi$}.
From this perspective, fringe phases tell us about asymmetric
information in an image and, as such, are unnecessary for
characterizing simple objects such as (idealized) stellar photospheres
or ellipsoidal distributions (such as flat inclined disks).

\subsection{What Is the Closure Phase?}
\begin{figure}[hbt]
\begin{center}
\includegraphics[clip,width=6in]{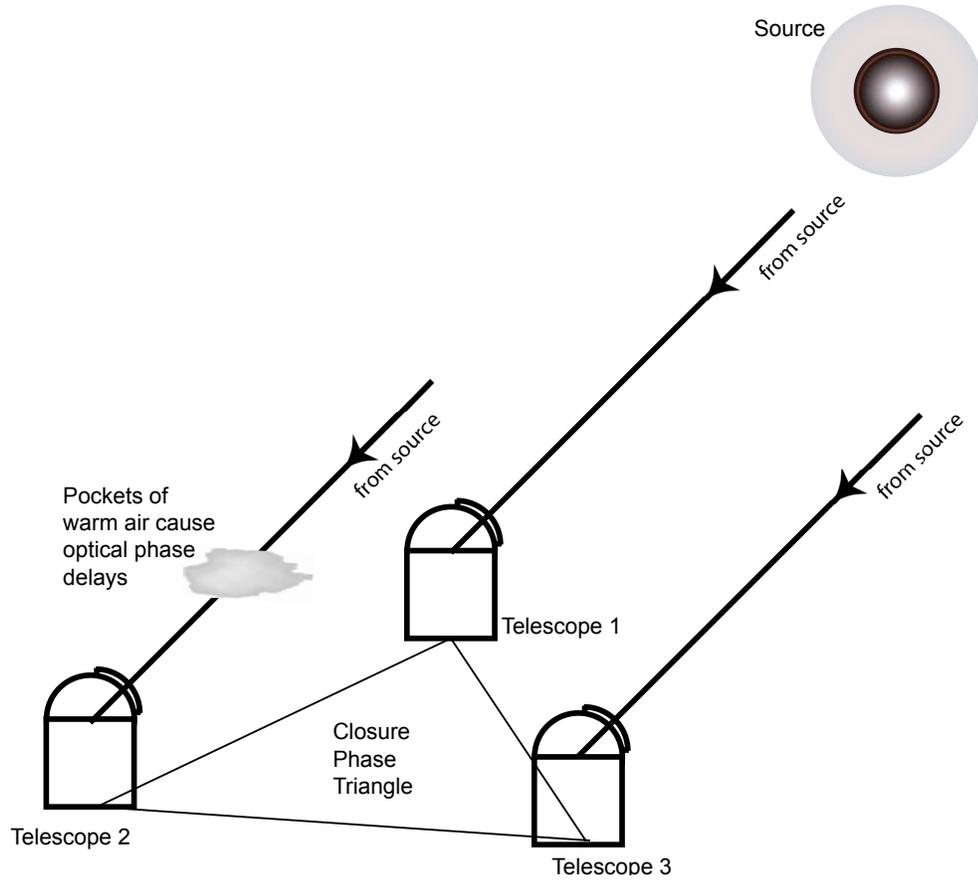}
\figcaption{This figure shows the telescope geometry for closure phase
measurement.  Turbulent air pockets cause optical path length variations above 
individual telescopes corrupting fringe phase measurements.  However by
summing fringe phases around a closed triangle, these phase errors 
cancel out resulting in a good interferometric observable, {\em the 
closure phase}.  
\label{fig_cpfig}}
\end{center}
\end{figure}

The closure phase method was developed by early radio
interferometrists \citep{jennison58} to calibrate phase drifts in
phase-coherent (heterodyne) detection systems, and it was first
applied to optical data in the context of aperture masking
\citep{haniff1987}.  The basic idea can be explained by considering a
three-telescope array as pictured in Figure~\ref{fig_cpfig}.
Consider that turbulent air pockets above aperture 2 will cause equal but opposite
effects on the two baselines connecting apertures 1-2 and apertures
2-3.  Thus, {\em summing up} phases around a triangle of 3 telescopes
cancels out all atmospheric disturbances within the ``closed''
triangle, resulting in a good observable, the closure phase, that is
independent of the phase fluctuations above the array of telescopes.

The following derivation mathematically expresses the qualitative
ideas of the last paragraph.  Consider the the closure phase
($\Phi^{\rm CP}_{ijk}$) for a triangle connecting telescopes $i$, $j$,
and $k$.

\begin{equation}
\Phi^{\rm CP}_{ijk} = \Phi^{\rm obs}_{ij} + \Phi^{\rm obs}_{jk} +\Phi^{\rm obs}_{ki}
\end{equation}
where, $ \Phi^{\rm obs}_{ij}$ here is the {\em observed} fringe phase
for baseline between telescopes $i$ and $j$.  The observed fringe
phase can be seen to be equal to the {\em intrinsic} phase $\Phi_{ij}$
(this is what we are most interested in) and the difference between
the time-varying atmospheric phase delays $\phi_i$ associated with
individual telescopes.  Thus,

\begin{equation}
 \Phi^{\rm obs}_{ij} =  \underbrace{\Phi_{ij}}_{\rm intrinsic} +
\underbrace{ (\phi_j - \phi_i) }_{\rm atmospheric}
\end{equation}

Putting these last two equations together we can prove the closure phase to 
be  independent of atmospheric phase delays:

\begin{eqnarray}
\Phi^{\rm CP}_{ijk} &=& \Phi^{\rm obs}_{ij} + \Phi^{\rm obs}_{jk} +\Phi^{\rm obs}_{ki}\\
                  &=& \Phi_{ij} +  (\phi_j - \phi_i) + \Phi_{jk} +  (\phi_k - \phi_j) + \Phi_{ki} +  (\phi_i - \phi_k) \\
                  &=&  \Phi_{ij} +\Phi_{jk} +\Phi_{ki}
\end{eqnarray}

In principle, images can be reconstructed using only visibilities and 
closure phases
\citep[related to the technique of ``self-calibration'' in
radio interferometry;][]{cw81} when additional constraints are imposed (such as
limited field-of-view and positivity).  A more pedagogical description
and derivation of the closure phase can be found in
\citet{monnier_mss} and \citet{monnier2003}.

\subsection{Important Properties of Closure Phases}
The most important property of closure phases follows directly
from the properties of the Fourier phases already discussed.
That is, a {\em centrosymmetric} (point-symmetric) image will always have
a closure phase of zero degrees\footnote{Actually, 180~degrees is also possible
but not relevant for our case of resolved structure around a dominant
point-like star.}.  Thus, detection of non-zero closure phase is a robust
sign of {\em skewed} emission, a definitive deviation from 
point-symmetry.   This remarkable and robust property of closure
phases is key for studies of young stellar objects, since model-independent
image reconstructions are today beyond our capabilities (due to limited
uv-coverage).

Another key point is that the interferometer must have sufficient
angular resolution to {\em resolve} the scale over which the emission is
skewed: an unresolved source is always point-symmetric.  Thus, 
one must match interferometer resolution to the appropriate angular scale.

While the detection of skewed emission is robust, quantifying the
degree of skew is very model-dependent.  When limited uv-coverage
forbids direct image reconstruction, one must carry-out model fitting.
Fortunately, disk theorists routinely produce models for fitting to
spectral energy distributions and the results of radiative transfer
models are suitable for comparing to interferometry data
\citep[e.g.,][]{malbet1995,dalessio1998,whitney2003,lachaume2003b,akeson2005b}.

\subsection{Skewed NIR disk emission from YSOs}
\label{skewednir}

In this paper, we compare measured closure phases with
physically-motivated models for the disk emission in order to quantify
the level of skewed emission present in our sample.  While future
closure phase surveys with larger telescopes should provide a rich and
detailed dataset for modeling, we show below that even the limited,
first-generation closure phases presented here can distinguish between
current competing disk  models.  Here, we contemplate two classes of disk models:
the DDN disk model and the ad hoc ``skewed ring'' model.
  
\subsubsection{DDN Models}
\label{ddn}
Dullemond, Dominik, \& Natta (2001, DDN) established the current
paradigm for Herbig disks \citep[and by extension, even new T~Tauri
disk models;][]{muz2003}.  As initially outlined by \citet{natta2001},
the star is surrounded by an optically-thin cavity with a gas-only
disk.  The dusty disk begins when the temperature drops below the dust
destruction temperature ($T\sim1500$K), at which point the midplane
becomes optically thick.  At the dust destruction radius the disk is
clearly quite hot and this ``puffs-up'' the inner rim.  The main
achievement of \citet{dullemond2001} was to couple the radiative
transfer to a (vertical) hydrostatic disk structure, finding a
physically self-consistent height for the inner rim and the outer disk
flaring \citep[which match on to traditional flared disk models,
e.g.][]{chiang1997}.

In the DDN paradigm, the frontally-illuminated inner wall is perfectly
vertical.  These authors comment that such a situation may not be
physically reasonable, and indeed a variety of effects may ``curve''
the inner rim (some of these are discussed in \S\ref{discussion}).  In
the case of a vertical wall, an observer will generally only see the
far side of the inner rim wall, since the front side will be blocked by
outer disk material. This scenario produces {\em maximally}-skewed disk
emission, boding well for closure phase studies of YSOs.

\begin{figure}[hbt]
\begin{center}
\includegraphics[angle=90,clip,width=6in]{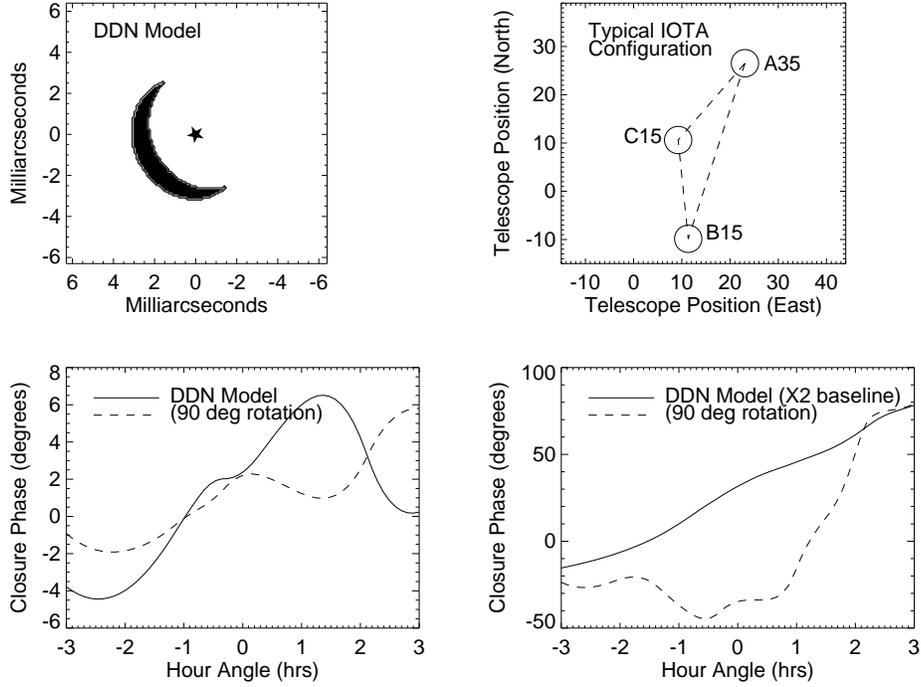}
\figcaption{This figure shows an example closure phase calculation for a
DDN disk model using IOTA.  (a) The top left panel shows the near-IR synthetic image
assuming an inner rim radius of 3~milliarcseconds, viewed at inclination $i=30\arcdeg$.
(b) The top right panel shows a typical observing configuration for the
IOTA 3-telescope interferometer.  (c) The bottom left panel shows the closure phase
as a function of hour angle, assuming object at declination $\delta=35\arcdeg$ 
(dashed line for the DDN model rotated $90\arcdeg$ 
on the sky).  (d). The bottom right panel shows the closure phase signal for
baselines twice as long, illustrating sensitivity with angular resolution.
\label{ddndisk}}
\end{center}
\end{figure}

Figure~\ref{ddndisk}a shows a synthetic image for the near-IR emission
from a DDN disk with 6~mas inner rim diameter, inclined to our
line-of-sight by 30$\arcdeg$.  The hot inner wall is assumed to have
uniform surface brightness and we neglect any near-IR emission or
scattering from the rest of the (cooler) disk.  Here we take the
half-height of the wall to be $\sim$0.3$\times$ inner radius
\citep[following][]{dullemond2001}.  As expected, the disk emission is
highly off-center -- {\em skewed} -- with respect to the star.

We can calculate the expected closure phase signal for this disk model
using the IOTA interferometer geometry shown in Figure~\ref{ddndisk}b,
a typical layout used for the observations in this paper.
Figure~\ref{ddndisk}c show the predicted closure
phase signal for the DDN model (and also a rotated version) for
$\lambda=1.65\mu$m.  Here we also assumed that 65\% of the
emission came from the dust while 35\% arises from the star (typical
decomposition for our targets). Figure~\ref{ddndisk}d shows how the
signal is much stronger if the interferometer has 2$\times$ longer
baselines -- clearly, the disk appears more ``skewed'' with increasing
angular resolution.

\subsubsection{ Generic ``Skewed Ring''  Model }
\label{skew}
Later in this paper, we will show that the DDN model predicts large
closure phases that are not observed.  In order to move beyond the
vertical wall assumption of the DDN model, we will explore the class
of ``skewed ring" models.  \citet{mm2002} discussed in detail the
merits of a ring model for fitting to the near-infrared visibilities
of YSOs, and this basic model has been used by most
current workers in this field \citep[e.g.,][]{rmg2001,eisner2004,akeson2005b}.  
The narrow range of temperatures over which dust emits at
1.65$\mu$m and the empirical results of \citet[][imaging of the
LkH$\alpha$101 disk]{lkha2001} both suggest the dust at the inner edge
of the YSO accretion disk will appear as a thin ring in the
near-infrared.  Here we adopt a Gaussian cross-section for the ring
emission, in order to approximate smoother (i.e. curved) inner rim
emission.

\begin{figure}[hbt]
\begin{center}
\includegraphics[angle=90,clip,width=6in]{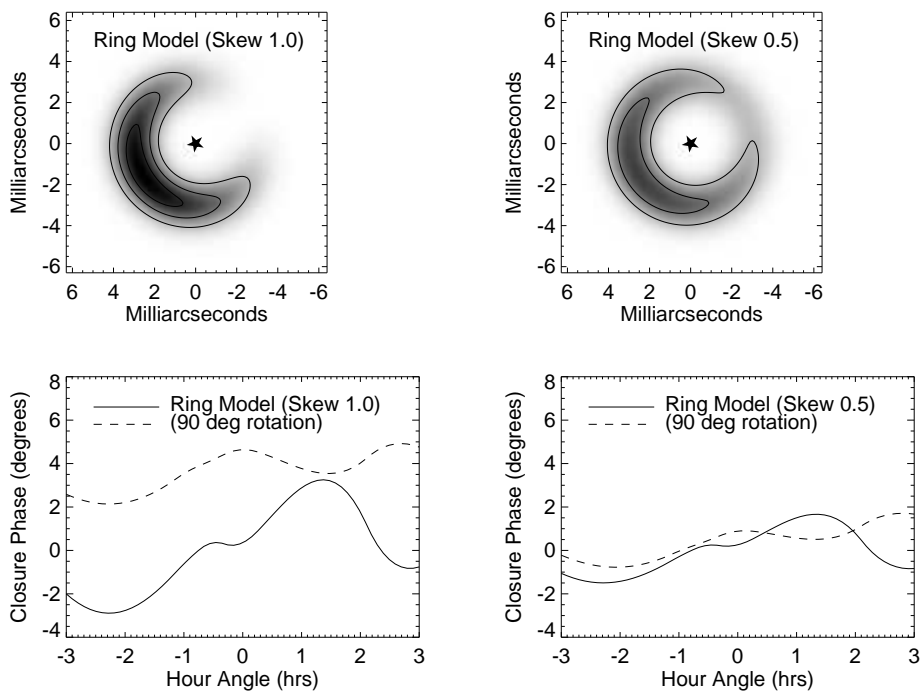}
\figcaption{This figure shows predicted IOTA closure phase signal for
examples of ``skewed ring'' models with similar geometry to the DDN model 
seen in Figure~\ref{ddndisk} (see \S\ref{skew} for model details).
(a) Top-left panel shows synthetic image for ring model with maximum skew $=1$.  (b)
Top-right panel shows synthetic image for ring model with intermediate skew $=0.5$.
(c) Bottom-left panel shows the closure phase result for skew $=1$ disk
(same IOTA observing  geometry as in Figure~\ref{ddndisk}b).  
(d) Bottom-right panel corresponds to ring model with skew $=0.5$.
\label{ringmods}}
\end{center}
\end{figure}

Here, we accept this ring model with an additional complication -- we
modulate the ring brightness as a function of azimuth by a sinusoid
(essentially, an $m=1$ mode).  This model was recently used for
interpretation of the AB~Aur disk by \citet{rmg2006}.
Figures~\ref{ringmods}a~\&~b show two different ring models with a ``skew"
of 1 and 0.5 (and ring thickness of 25\% using Gaussian profile).  The
``skew" refers to the amplitude of the modulation in the ring
brightness, thus a skew of 0 is centrosymmetric, a skew of 0.5 has a
3-to-1 brightness contrast between the brightest and faintest
(diametrically-opposed) portions of the ring, and a skew of 1 has a
contrast formally of $\infty$.  You will note that the skew 0.5 image
is remarkably similar to the image of LkH$\alpha$101 reported by
\citet{lkha2001}.

Figures~\ref{ringmods}cd show the observable closure phases for these
models using the same observing geometry introduced for
Figure~\ref{ddndisk}.  The closure phase signal for the skew$=1$ model
is smaller than the DDN model, as expected since the emission is far
less skewed.  Also, we see that the skew$=0.5$ model presents even
smaller closure phases for the same observing conditions.  These
calculations illustrate clearly how precision closure phase
measurements of YSOs by today's interferometers can determine the
``skewness'' of the near-IR emission, discriminating between competing
disk models (see \S\ref{analysis}).

If the skewness in disk emission is a result of a disk viewed at
intermediate inclination, we might plausibly expect the skewed ring
to be elliptical and not circular.  This level of detail is beyond the
current constraints from the interferometer data and would introduce
another free parameter.  Thus, we will not consider this
reasonable extension to the ``skewed ring'' model for the majority of our
targets (we do introduce such a sophisticated model for the
special case of HD~45677 in \S\ref{hd45677cp}).

\section{Analysis of Closure Phases}
\label{analysis}
Except for AB~Aur \citep{rmg2006}, none of our targets have previously been
observed using closure phases and we will concentrate our analysis of
this section on this observable.  We find 8 of the 14 sample targets
have closure phases consistent with zero within measurement errors.
Only AB~Aur, HD~45677, MWC~297, MWC~361-A, MWC~614, MWC~1080 show
statistically-significant non-zero closure phases (at 2-$\sigma$ level).

In order to interpret this result, we must consider the angular
resolution of the IOTA interferometer and disk sizes of our targets
(measured in \S\ref{section:results1}), since many of our disks are
only marginally-resolved.  Recall from \S\ref{cpintro} that closure
phases encode deviations from centrosymmetry in an extended object,
closely related to the ''skewness" of the emission.  Perhaps best
explained in \citet{lachaume2003}, marginally resolved objects
naturally possess small closure phases even if the inherent emission
is extremely asymmetric.  An intuitive explanation of this is to
consider that any ``unresolved" source must be considered to be
centrosymmetric.

Thus, our first priority is to determine whether the
finding of small closure phases
reflects a true lack of skewness in the YSO disk emission, or rather
results from from inadequate angular resolution.  To further this
investigation, we have collected the maximum closure phase observed
for each target in Table~\ref{cpresults}, along with the disk size
measured in \S\ref{section:results1} in units of the fringe spacing
($\lambda/B$) of the longest IOTA baseline in the appropriate triangle.

We confront the specific predictions of the DDN model in the next section.

\subsection{Confronting DDN Models}

In order to extract quantitative information from the
near-zero closure phases of most stars in our sample, we will first 
explore the closure phase predictions of DDN \citep{dullemond2001} models,
described in \S\ref{ddn}.

\begin{figure}[hbt]
\begin{center}
\includegraphics[angle=90,clip,width=6in]{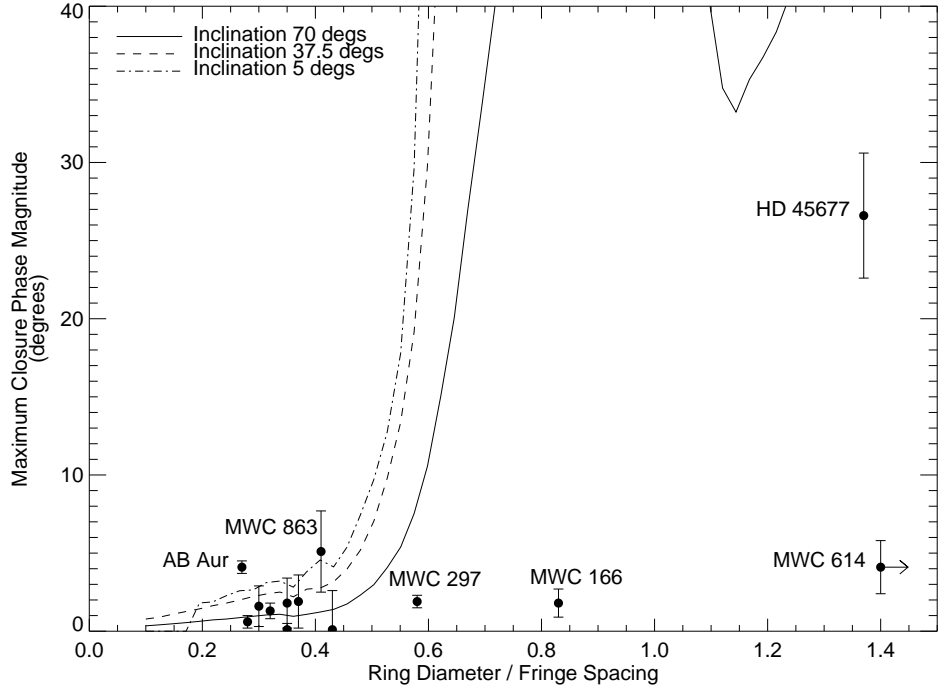}
\figcaption{In this figure we compare the maximum
predicted closure phases for DDN models as a function of disk inclination angle and
angular resolution of interferometer.  
For well-resolved disks, the closure phase signals for DDN models are
expected to be huge but are not observed in any targets of our sample. 
A few of the most notable targets are labeled. 
\label{cpresults_ddn}}
\end{center}
\end{figure}

In order to explore parameter space, we generated closure phase
predictions for DDN models with a range of inner rim diameters and
viewing angles.  In each case, we assumed the disk made up 65\% of the
emission at H-band with the remaining 35\% arising from the
(unresolved) star, a typical case for our target sample.  We then
generated synthetic closure phases for the actual observing triangles
used by IOTA.  The main results of this parameter study are shown in
Figure~\ref{cpresults_ddn}.

In Figure~\ref{cpresults_ddn}, we plot the {\em maximum observable}
IOTA closure phase for DDN disks for three different inclinations as a
function of the ring diameter.  The ring diameter is expressed here in
units of the fringe spacing $\lambda/B$ of the longest baseline in a
given IOTA closure triangle.  For each skewed ring we sampled all
possible position angles on the sky and here plot the maximum value of
the closure phase magnitude. While we typically do not know the sky
orientation of a given object, we expect our sample of 14~targets to
sample random position angles on the sky.  

This procedure quantifies the qualitative results from the DDN
examples in Figure~\ref{ddndisk}, that well-resolved DDN disks have
large closure phase signals.  In Figure~\ref{cpresults_ddn}, we can
then compare the synthetic curves to the maximum closure phases
actually observed in our survey.  To do so only requires us to
account for the differing sizes of each target from
\S\ref{section:results1}.  We exclude MWC~361-A from this analysis since
the large closure phases for this target likely come from the binary
companion, not skewness in the disk around the primary star.

Based on this comparison of DDN model predictions with our closure phase
dataset, we can draw 2 primary conclusions.

\begin{itemize}
\item{The most well-resolved disk sources have smaller closure phases than
the DDN model calculations, though 
usually non-zero.  This indicates some modest skewness in the disk emission
but is incompatible with the large skew signal predicted by DDN models.}
\item{When the ring diameter is less than the $\sim\frac{1}{2}$ fringe
    spacing of the longest baseline in a triangle, the closure phase
    signal is strongly suppressed.  This applies to 
   approximately half of our  target sample.  We will require approximately 2-3$\times$ better angular resolution to fully
characterize disk skewness in our sample (corresponding to 70-100m baselines at H
band or 100-150m baselines at K band).}
\end{itemize}

For any given target, the putative disk skewness might be orientated
along a position angle not sampled by the IOTA interferometer, which
is most sensitive to skew in the N-S direction (due to non-uniform
uv-coverage).  Thus, even highly skewed disk emission can be
``hidden'' from the interferometer due to unfavorable observing
geometry.  However, the small observable closure phases for the most
resolved disks (MWC~297, MWC~166, MWC~614, MWC~480, and RY~Tau), when
taken as a whole, argue strongly against the vertical wall of the inner
rim put forward in the DDN model, since this geometry always
predicts strong skew in disks viewed from all inclination angles.
Note that the DDN model (with vertical inner wall) also predicts
highly {\em elongated} disk emission (see Figure~\ref{ddndisk}) even
for nearly face-on disks, inconsistent with the \viss results of
\citet{eisner2004}.  Note that AB~Aur uniquely appears to show definite skewed
emission {\em in its halo} (beyond the inner rim), 
further discussed in \citet{rmg2006}.

\subsection{Confronting ``Skewed Ring'' Models}

\begin{figure}[hbt]
\begin{center}
\includegraphics[angle=90,clip,width=6in]{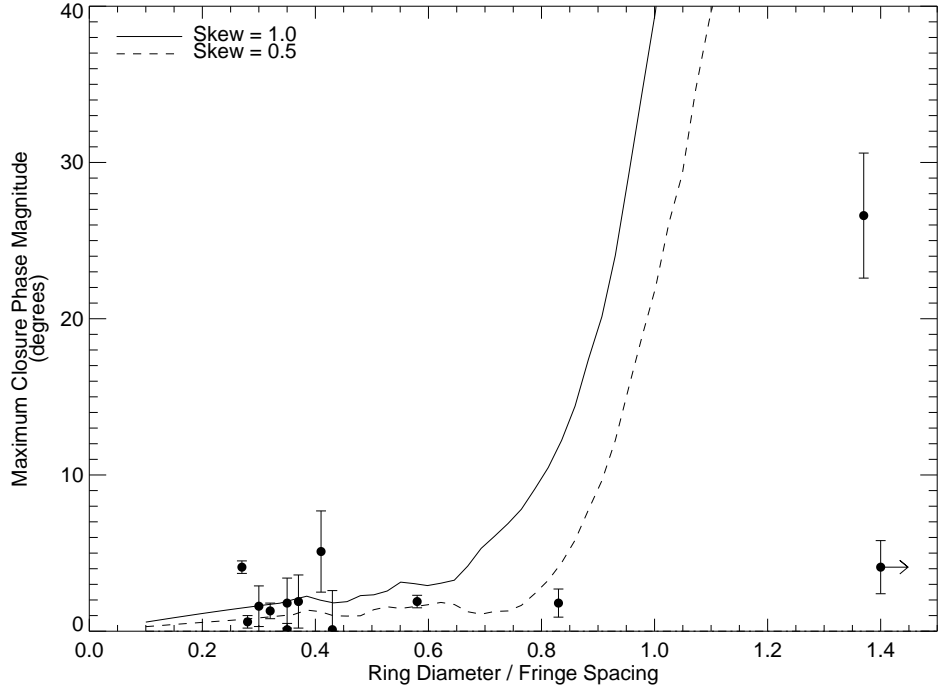}
\figcaption{In this figure we compare the maximum
predicted closure phases for skewed ring models as a function of the skew amount and
angular resolution of interferometer.  Note that the
line for skew $=$0 is identically zero due to centrosymmetry.
Note that most systems are not resolved enough to
constrain the skewed models, calling for longer-baseline interferometers such as
PTI, CHARA, and VLTI.
\label{cpresults_skew}}
\end{center}
\end{figure}

The analysis of the last section was also carried out using the ``skewed
ring'' model described in \S\ref{skew}.  The results of this study,
along with the observed closure phases, are plotted in
Figure~\ref{cpresults_skew}.  While skew$=1$ disks do generally
overpredict the closure phase magnitude, our data can not clearly
distinguish between the models with varying ``skew'' factors.
Although ad hoc, the skewed ring models are physically motivated and
can currently explain the observed visibility and closure phase data
in our sample, including the the imaging data for LkH$\alpha$~101 and
the closure phase data of HD~45677 (taking ellipticity into account; see \S\ref{hd45677cp}). 
While it
is critical to develop {\em physical} models for the disk emission,
the introduction of the skewed ring model offers a new and useful tool
for describing the emission independent of the underlying
physical models.

For some disks, it should be possible to do more detailed model
fitting by including inclination and position angle information known
from CO data or scattered light images (e.g., AB~Aur, MWC~275). In
these cases, the IOTA closure phases can play a more definitive role
in testing specific models.  However, in general, longer baseline
closure phases and more uniform Fourier coverage will be needed to
refine future models of the inner rim geometry.

\subsection{Discussion}
\label{discussion}

\subsubsection{Implications on inner rim geometry}

The most significant result from this paper is the report of few
skewed disks around Herbig Ae/Be stars.  This is robustly shown by the
uniformly small closure phases found in our sample of YSOs.  When the
sample is viewed as a whole, the small closure phases have been shown
to be incompatible with generic predictions of the original DDN disk
models; these models possess vertical walls at the inner edge of the
accretion disk introducing strongly-skewed emission at all viewing
angles.  Note that this result does not invalidate the overall success
and applicability of the DDN model which mostly focuses on other
aspects of the disk structure.  While we have shown that
``skewed ring'' models are better suited for explaining the small observed
closure phases, we also demonstrate that detailed determinations of ``skew''
required longer baselines than possible with IOTA.

%%%%%We attempted to further quantify the
%level of skew seen in our sample using the ``skewed ring'' models, but
%were unable to make further progress due to the small sample size and
%the fact that most YSO disks were not well-resolved by IOTA.

In the context of today's disk models, our data lend support to inner disk
models like that of \citet{isella2005} which includes a {\em curved} inner rim.
In this case, the inner rim curves away from the midplane  
due to the expected pressure-dependence of the dust
evaporation temperature.  We plan to explore quantitatively the expected
closure phase signals of this model with future radiative transfer
work, but this is beyond the scope of this paper.

Note that there are other mechanisms that could cause the inner rim to
``curve.''  One new idea we suggest is a consequence of dust settling
and growth which leads to larger dust grains in the midplane than in
the upper layers of the disk. Since large grains better radiate in the
infrared, they are able to exist closer to the star.  Smaller grains
in the upper layers have difficulty cooling and thus evaporate more
easily.  Thus, the inner rim can be highly curved if the average dust
size varies significantly as a function of disk scale height in the
inner rim. This process can cause a much greater curvature in the
inner rim compared to the effects of gas pressure on the evaporation
temperature.  Indeed, other recent work have found evidence for dust
settling and growth in YSO disks \citep[e.g.,][]{rettig2005,
  duchene2004}, and the implications on the inner rim 
structure have only just begun to be explored  \citep[e.g.,][]{tannirkulam2005}.

Understanding the geometry of the inner disk is not just important for
understanding the near-IR emission.  \citet{vanboekel2005} showed that
20-30\% of the mid-IR emission can come directly from the inner rim.
Given the intriguing changes in the silicate feature observed using
the MIDI instrument on VLTI \citep{vanboekel2004}, it is critical to understand
how the dust properties of the rim might be different than in the
surface layers of the disk.  In addition, high spectral resolution CO
observations of near-IR fundamental and first-overtone lines \citep{najita2003,brittain2003,blake2004}
require temperature and density modeling of the inner AU of Herbig
disks.  Currently, these models have not incorporated the puffed inner
wall into their calculations, an improvement urgently needed in order to
correctly interpret the potentially powerful kinematic and temperature
information available.

%replace rettig2005 AAS with rettig2006 ApJ when available.

\subsubsection{The ``Halo'' Phenomenon}

By using three telescopes, our survey has measured visibilities on a
larger range of scales than previous work using only two telescope
interferometers.  We find evidence for intermediate- to large-scale
``halos'' around AB~Aur, MWC~275, MWC~297, and RY~Tau.  For our purposes
here, we define halo emission to be any extended emission on scales
between 10-1000~milli-arcseconds making up 5-20\% of the near-IR flux.
Similar extended halos have been reported previously (albeit on
0.5-2.0~arcsecond scales) around some T Tauris and Herbigs using
infrared speckle interferometry \citep{leinert1993,leinert2001} and
even on smaller scales probed by lunar occultation \citep[e.g., case
of DG~Tau;][]{leinert1991, chen1992}. Similarly-sized extended
emission has recently been reported around a sample of FU~Ori objects
using the Keck Interferometer \citep{rmg2006b}.

Existence of these halos is not well-understood theoretically, but
could arise from diffuse scattered light at large (5-100 AU) radii 
\citep[as recently discussed by][]{akeson2005}.
from the central star or even from localized thermal emission within a
few AU.  The origin of the halo material is unclear, but plausibly
could be from an infalling remnant envelope, dust entrained in the
stellar wind/outflow, or (in the case of FU Ori) ejected gas \& dust
from an earlier outburst.  The connection to the hybrid halo+disk
model of \citet{vinkovic2006} is ambiguous and our findings should motivate
further work on such multi-component models.  In the special case of
AB~Aur \citep{rmg2006}, we detected non-zero closure phases from the
``halo'' component, leading us to conclude the source of the emission
was localized and within 1-4~AU of the star.  For the other ``halos''
reported in this paper, we can only roughly constrain the location of
the emission (10-500 milliarcseconds); further observations with {\em
  short baselines} will be required to elucidate the nature of
``halos'' around Herbig Ae/Be stars.

Recently, \citet{baines2006} report interesting ``spectro-astrometry''
observations of many Herbig stars finding evidence for possible
binaries around many targets (including AB~Aur, HD~45677, v1295~Aql,
MWC~361-A of this sample).  While we caution that a few of our ``halo''
detections may turn out to be due to very close binary companions
(10-500~mas),  most stars in our sample have been surveyed for companions
using adaptive optics down to about 100~mas.\citep[e.g., AB Aur,
MWC~297, v1295~Aql;][]{eisner2004}.  Similarly, the spectro-astrometry
data can misinterpret halo emission as evidence for binarity if the
extended emission itself is skewed \citep[as has been shown for
AB~Aur;][]{rmg2006}. Note that scattered disk/halo light can only affect
spectro-astrometry results if it has a substantially different H$\alpha$
spectrum than what we see from the star. This could happen perhaps in two ways.  The
H$\alpha$ emission from the star is probably not uniformly distributed,
but more concentrated near polar regions and will also have a different
dynamical signature/line profile -- thus, the observer will
see different H$\alpha$ than the disk midplane which then scatters the
light into our line-of-sight.  Second, stellar light must traverse long paths just
above the disk midplane before scattering into our line of sight, thus
there could be significant H$\alpha$ {\em absorption} (depending on
ionization state). These effects deserve more study before settling on an
interpretation of the spectro-astrometry data.

\section{Special Treatment for HD~45677 and MWC~361}

\subsection{Parametric Imaging of the HD~45677 Disk}
\label{hd45677cp}

One exception to the general pattern of small closure phases is
HD~45677, which is both heavily resolved, elongated, and shows strong
closure phases.  We already fit a symmetric ring and Gaussian to the
model in \S\ref{section:results1} and Figure~\ref{sedvisHD45677}, and
noted the large fit residuals due to elongated and skewed structures
evident in the raw \viss and closure phase data.

\begin{figure}[hbt]
\begin{center}
\includegraphics[angle=90,clip,width=6in]{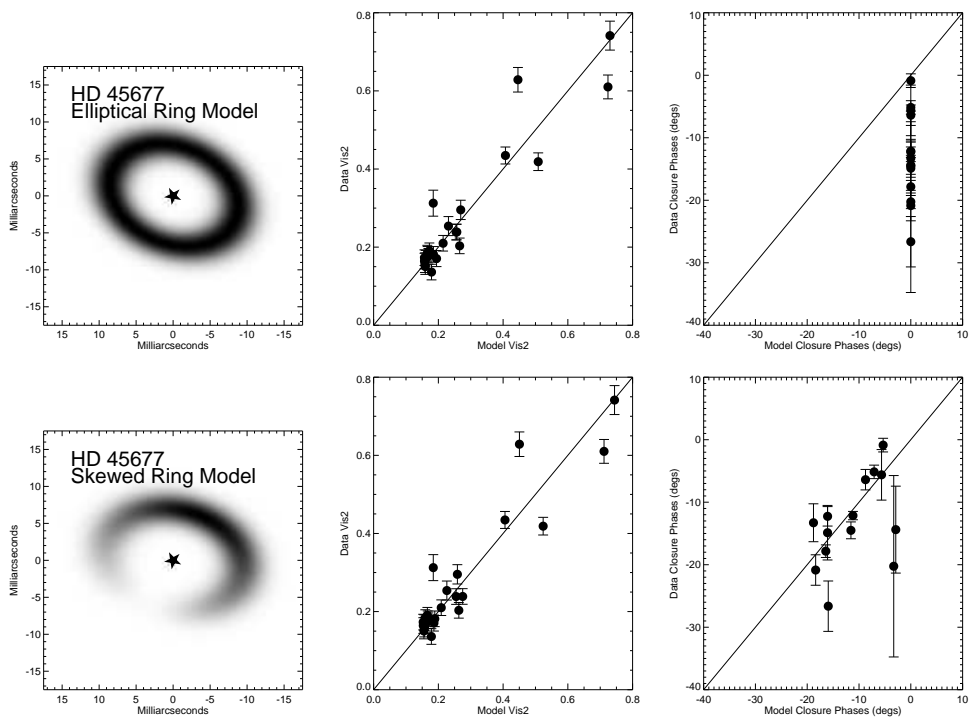}
\figcaption{Top-left panel shows the best-fit Elliptical Ring Model to the
IOTA3 data (North is up, East is left).  Top-middle panel compares the observed visibility data with 
the model data.  Top-right panel compares the observed closure phases with the
model closure phases.  Since this model is constrained to be centrosymmetric,
all model closure phases are identically zero and thus a good fit is impossible.
The bottom panels present the best fit ``skewed elliptical ring model.''
The closure phases are fairly well-explained by a strong north-west skew.
All parameters of these fits are presented in Table~\ref{hd45677results}.
Some data points are not well-explained by these models suggesting 
a patchier dust distribution.
\label{modsHD45677}}
\end{center}
\end{figure}

First we fit the \viss data with an elongated ring model, and the
results of this fit are shown in the top panels of
Figure~\ref{modsHD45677}.  The direction of elongation (PA
70$\arcdeg$) is similar to the observed (visible) polarization angle
of 60-80$\arcdeg$ \citep{coyne1976}.  However, as expected, this model
can not explain the relatively large observed closure phases
that must arise from skewed emission.  The results of this fit are
tabulated in Table~\ref{hd45677results}.

In order to explore this, we allowed the elongated ring to be skewed,
following the same method as for the skewed ring model in
\S\ref{skew}.  That is, the elongated ring emission is modulated as a
function of azimuth.  The best fit parameters are found in
Table~\ref{hd45677results} and the synthetic image is shown in
Figure~\ref{modsHD45677} (bottom panels).  Interestingly, the skew
angle (position angle from central star with greatest skewed brightness)
is $\sim$90~$\arcdeg$ different than the major axis of the
elongated ring, a geometry that might be expected for a flared disk
\citep[e.g.,][]{malbet2001}.  Given our parametric image
and the direction of the skew, we would predict that scattering by
dust would produce a net linear polarization of PA $\sim$70$\arcdeg$,
in agreement with observations.  The $\chi^2$ fit to
the \viss is poor at the shortest baselines suggesting additional complexity
missing from our model.

The skewed ring model presented here can be viewed as the first
attempt to do ``parametric'' imaging of YSO disks using long baseline
interferometry. While the number of degrees of freedom in our model is
still relatively small, we are reconstructing fairly complicated image
characteristics.  Similar quality ``images'' should be possible for
the rest of our target sample in the coming years.

Lastly, we want to comment that this ring image is qualitatively similar to the
expected results outlined in previous papers \citep{coyne1976,oudmaijer1999}.  Based on
the relatively cool infrared excess and the linear polarization angle,
a ring-like, elongated morphology was predicted and the position
angles found here are entirely consistent with the polarization
signatures.  Combining infrared polarimetry with interferometry could
reveal unique aspects of the dust in this system
\citep[e.g.,][]{ireland2005}.

We believe that our data support the idea that HD~45677 is not a {\em
  bona fide} Young Stellar Object, but rather a more evolved system
akin to CI~Cam \citep[e.g.,][]{mirosh2002}. The disk size is too large to be
a typical YSO system (unless we are observing a unique transition
object). Perhaps a binary interaction in the recent past created the
disk-like dust distribution around this source \citep[e.g.,][and references therein]{mirosh2005}.
Studies of this dust shell may shed light on the mysterious process of
dust production around some main-sequence Be stars.

\subsection{Astrometric Orbit for the Young Binary MWC~361-A}
\label{mwc361orbit}

The first definitive detection of the MWC~361-A close binary was
reported by \citet{rmg2001} using interferometry. As previously
already mentioned, spectroscopic evidence has also been accumulating
and \citet{pogodin2004} recently reported a preliminary
single-lined radial velocity (RV) orbit of the primary with a period
of 1341~days.

There are few simultaneous radial velocity and astrometric orbits for
young star binaries \citep[e.g.,
HD~98800-B;][]{boden2005}\footnote{Eclipsing binary systems are also 
  powerful tools for determining stellar masses
  \citep[e.g.,][]{stassun2004}.}  These data are particular valuable
because they directly constrain the stellar masses and sometimes the
system distance, invaluable ``ground truth'' tests for theories of
stellar interiors and evolution.

We report five new epochs of \viss and closure phase data for MWC~361-A
in this paper.  All
these epochs have been fitted for a binary system, with the additional
complication that the primary is clearly partially resolved.  For the treatment
in this paper, we have assumed the fainter secondary
star is completely unresolved as expected for the size-luminosity relations of
Herbig stars.   We also have independently re-fit the
two epochs of data from
\citet{rmg2001} in order to extend our time coverage over $\sim$5 years, 
more than 1.5 orbital periods.  

Table~\ref{binarypositions} shows our best-fit positions for the
secondary component relative to the primary for each of the seven
epochs.  Furthermore, our fitting found a brightness ratio of
6.5$\pm$0.5 (primary/secondary) at H band and a primary uniform disk
diameter of 3.6$\pm$0.5 mas (we analyzed the primary disk size in
\S\ref{section:results1} and Figure~\ref{sedvisMWC361}).  There was
some evidence for variations in the flux ratio with time, as expected
for young stars and previously observed for the Z~CMa system
\citep[e.g.,][]{rmg2002}. For about half the epochs (1998 June, 2003
November, 2004 June, 2005 June) we were able to uniquely constrain all
the free parameters.  For the remaining, we had to fix the
brightness ratio and size of the components to yield a good positions
(due to more limited datasets).
In a few cases, we had to use {\em a priori} information concerning
the likely binary position to break position degeneracies. For
instance, the 1998 September data comes only from one night and
multiple solutions exist -- however, only one solution is sufficiently
close to the unambiguous result of 1998 June to be considered
plausible and consistent with the other epochs.

\begin{figure}[hbt]a
\begin{center}
\includegraphics[angle=90,clip,width=6in]{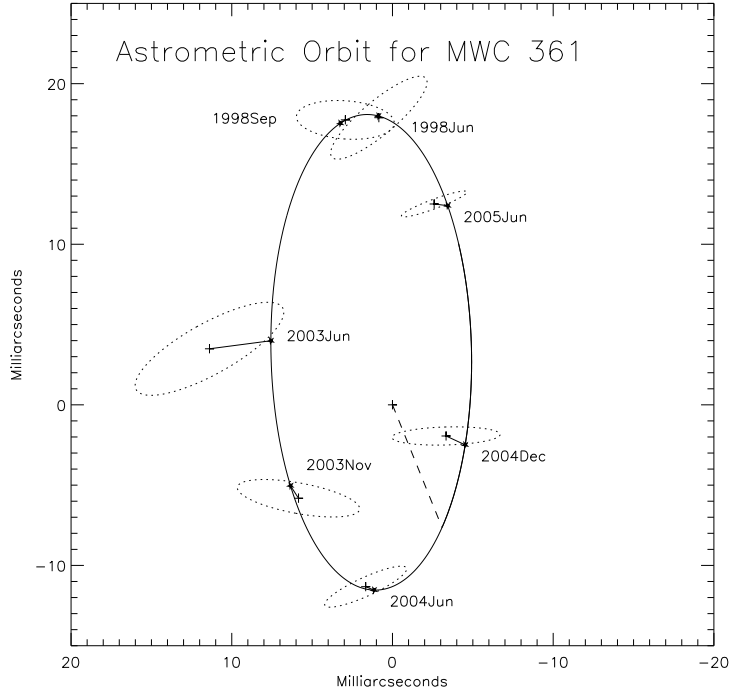}
\figcaption{This figure shows the astrometric orbit for MWC~361-A (solid line). The 
observed locations of the secondary component (relative to primary) are marked with
error ellipses (dotted lines with crosses) for seven different epochs.  The dashed line
connects the locations of the primary and secondary star at periastron passage.
\label{orbitMWC361}}
\end{center}
\end{figure}

Figure~\ref{orbitMWC361} shows error ellipses for the
position of the secondary relative to the primary for all epochs.  The
position predictions for our best-fit orbit are also shown and the
agreement is excellent.  The orbital parameters (and conservative
error estimates) are contained in Table~\ref{binaryparams} and
compared to the orbital elements of \citet{pogodin2004}.  All orbital
elements agree within observational uncertainties.  A future paper
will fit the \viss and closure phase data directly (along with radial
velocity data) using the orbital elements \citep[as recently done
using IOTA data for $\lambda$~Vir by][]{zhao2006}. This analysis will
yield a joint orbital solution with significantly-smaller parameter
errors, however this analysis is beyond the scope of this paper.

With the combined RV and astrometric orbital solution, we can
determine physical characteristics of the MWC~361-A system.  We
had the Hipparcos intermediate data re-analyzed using the new orbital
parameters in an attempt to improve the parallax and constrain the
photocenter orbit \citep[in the method of][]{pourbaix2000}.  The new
result $\pi = $2.76$\pm$0.68 mas (d$=360^{+120}_{-70}$~pc) varies
little from the original catalog result ($\pi =
$2.33$\pm$0.62~mas;d$=430^{+160}_{-90}$~pc).  Using the orbital period
(3.74$\pm$0.06~yrs; weighted-average between the independent RV and
astrometric determinations), and (angular) semi-major axis
(15.14$\pm$0.70~mas), we can calculate the 
total system mass $M_1+M_2 = 10.4^{+20.5}_{-5.9}$
using the original Hipparcos measurement (these errors represent 1-$\sigma$ confidence 
intervals using a
Monte Carlo method for error propagation).
Unfortunately, while these preliminary
masses are consistent with our expectations for a binary system of
early B stars, the large distance errors propagate into large mass
uncertainties, rendering these estimates useless for constraining any
models.  It will be crucial to extract a {\em double-lined} radial velocity orbit in
order to determine an {\em orbital} parallax, resulting in accurate and precise
mass estimates for both components.  

We note that \citet{pogodin2004} measured the mass function $f(M_2$)
for the secondary from the single RV orbit.  This result ($f(M_2) =
0.175\pm0.035$) was interpreted using an assumed inclination
$i=70\arcdeg$, consistent with our measurement of $i= 65\pm8 \arcdeg$.
Thus, we can concur with their analysis that MWC~361-A appears to be a
close binary consisting of an early Be star, likely hosting the
resolved disk of \S\ref{section:results1}, and a lower-mass, late Be
star as secondary.

The periastron distance of the MWC~361-A components is only $\sim$4~AU
(due to somewhat eccentric $e\sim0.3$ orbit), thus the accretion disk
around the MWC~361-A primary should be severely truncated
\citep[perhaps this relates to recent difficulty in fitting the 
far-infrared
SED,][]{elia2004}.  Certainly, MWC~361-A provides a fascinating laboratory for
studying binary star formation and is suitable for future studies by
longer baseline interferometers that can both measure orbital
parameters at high accuracy but also  image the accretion disks
around both stellar components.  It is extremely rare to find young
stars of such early spectral types in a binary system, and MWC~361-A
promises to be a new anchor point for models of stellar interiors and
evolution, as well as for disk evolution in binary systems.

\section{Conclusions}
\label{sectionc:conclusions}

Here we have presented the results of the first closure phase survey
of Young Stellar Objects using an infrared interferometer.  We find
most YSO targets show no signs of skewness in their emission on the scale
of $\sim$4 milliarcseconds.  Our data is incompatible with vertical
wall models of the puffed-up inner rim \citep[e.g.,][]{dullemond2001}.
A ``skewed ring'' model was developed in order to quantify our results
and the observed small closure phases favor inner dust rims which are
smoothly curved \citep[e.g.,][]{isella2005}.  At least a crude
understanding of the inner AU of YSO disks is crucial for interpreting
a host of other YSO observations, including spectrally-resolved CO
data, spectro-astrometry, mid-IR sizes and spatially-resolved silicate
feature minerology.  Baselines significantly longer than the maximum
IOTA length of 38\,m (i.e., from CHARA and VLTI) will be required to
investigate this further for specific sources.

An independent and complete reanalysis of new H band \viss data found
fresh evidence for extended ``halos'' of emission with about 5-10\% of
the emission for a few targets \citep[see also][]{leinert2001}.  Halos
appear to be quite common in many young disk systems (T~Tauri, Herbig,
FU~Ori), although the physical origin of this large-scale emission
appears, as of yet, not understood.  These new data may re-open the debate
concerning the relative importance of disks, halos, envelopes, localized thermal
emission, and the applicability of hybrid models.  

We also took the first steps towards imaging young stellar objects by
studying the unusual system HD~45677, finding this Be star to be
surrounded by an elongated and highly-skewed dust ring.  We agree with
earlier suggestions that HD~45677 may not be a young object, but rather
a member of the enigmatic class of main sequence hot stars with large
IR excess.

Lastly, we calculated the first astrometric orbit for the short-period
binary MWC~361-A, finding good agreement with recent spectroscopic work. We found
mass estimates in line with expectations and discuss the exciting science 
potential from future studies of this unique system.

\acknowledgments {JDM thanks D. Pourbaix for re-analyzing the
  Hipparcos parallax of MWC~361-A using the new orbital parameters and
  A. Tannirkulam for photometry from the MDM Observatory.  The authors
  gratefully acknowledge support from the Smithsonian Astrophysical
  Observatory, NASA (for third telescope development and NASA
  NNG05G1180G), the National Science Foundation (AST-0138303,
  AST-0352723), and the Jet Propulsion Laboratory (JPL awards 1236050
  \& 1248252).  EP was partially supported by a SAO Predoctoral
  fellowship, JDM by a Harvard-Smithsonian CfA fellowship, and RM-G
  and J-PB were partially supported through NASA Michelson
  Postdoctoral Fellowships.  The IONIC3 instrument has been developed
  by LAOG and LETI in the context of the IONIC collaboration (LAOG,
  IMEP, LETI). The IONIC project is funded by the CNRS (France) and
  CNES (France). This research has made use of the SIMBAD database,
  operated at CDS, Strasbourg, France, and NASA's Astrophysics Data
  System Abstract Service.  This publication makes use of data
  products from the Two Micron All Sky Survey, which is a joint
  project of the University of Massachusetts and the Infrared
  Processing and Analysis Center/California Institute of Technology,
  funded by the National Aeronautics and Space Administration and the
  National Science Foundation.  This work has made use of services
  produced by the Michelson Science Center at the California Institute
  of Technology.  }

\bibliographystyle{apj}
\bibliography{apj-jour,Review,Review2,Thesis,WR140,HerbigSizes,iKeck,RX_Boo,IONIC3}

%% Generally speaking, only the figure captions, and not the figures
%% themselves, are included in electronic manuscript submissions.
%% Use \figcaption to format your figure captions. They should begin on a
%% new page.

\clearpage

\begin{deluxetable}{lllllllll}
\rotate
%\tabletypesize{\scriptsize}
\tabletypesize{\tiny}
\tablecaption{Basic Properties of Targets\label{targets}}
%\tablewidth{0pt}
\tablehead{
\colhead{Target} & \colhead{RA (J2000)} & \colhead{Dec (J2000)} &
\colhead{V} & \colhead{H} &
\colhead{Spectral} & \colhead{Distance} &\colhead{Adopted} & \colhead{Photometry}\\
\colhead{Names} & & &\colhead{mag\tablenotemark{a}} & \colhead{mag\tablenotemark{a}} & \colhead{Type}  & \colhead{(pc)} & \colhead{Luminosity (L$_\odot$)} & \colhead{References}
}
\startdata
RY~Tau & 04 21 57.40 & +28 26 35.5 & 10.2 & 6.1 & F8IIIe (1) & 142$\pm$14 (2) & 5$\pm$2 (3)   & 4, 5, 6, 7 \\
MWC~480, HD~31648 & 04 58 46.26 & +29 50 37.1 & 7.7 & 6.3 & A5V (1) & 131$^{+24}_{-18}$ (8,24) & 15$\pm$5 (3) & 5, 6, 9, 10 \\
AB~Aur   & 04 59 41.53 & +40 50 09.7 & 7.1 & 5.1 & A1 (11) & 144$^{+23}_{-17}$ (12) & 70$\pm$20 (3) & 4, 5,  6, 7, 9, 10, 13, 14, 15, 16 \\
HD~45677\tablenotemark{b}, MWC~142   & 06 28 17.42 & -13 03 11.0 & 8.1 & 6.3 & B2 (17) & 1000$\pm$500 (17) & 14000$\pm$7000 (3)  & 4, 18 \\ 
MWC~166, HD~53367 & 07 04 25.52 &-10 27 15.7 & 7.0 & 6.2 & B0IVe (19) & 1150 (20)  &  100000$\pm$50000 (3) & 4, 5, 15, 18 \\ 
HD~144432, HIP 78943 & 16 06 57.95 & -27 43 09.4 & 8.2 & 6.5 & A9IVev (1) & 145 (21)  &  15$\pm$4.0 (3) &  4, 5,  15, 22,  23  \\
MWC~863, HD~150193 & 16 40 17.92 & -23 53 45.2 & 8.9 & 6.2  & A2IVe (1) & 150$^{+50}_{-30}$ (8,24) &  20$\pm$10 (3) &  4, 5, 6, 9, 14  \\  
MWC~275, HD~163296 & 17 56 21.29 &-21 57 21.8 & 6.9 & 5.5 & A1Vepv (1) & 122$^{+17}_{-13}$ (8,24) & 40$\pm$8 (3) &  4, 6, 9, 16, 22 \\
MWC~297, NZ~Ser & 18 27 39.53 & -03 49 52.0 & 12.3 & 4.4 & B1Ve (1) & 250 (25) & 33000$\pm$13000 (3)  &  4, 10, 14, 22 \\
MWC~614, HD~179218 & 19 11 11.24 & +15 47 15.6 & 7.2 & 6.6 & A0V (1) & 240$^{+70}_{-40}$ (8,24) & 100$\pm$35 (3) &  4, 5, 6, 15, 22,  26 \\
v1295~Aql, HD~190073 & 20 03 02.51 & +05 44 16.7 & 7.8 & 6.6 & A2IVev (1) & $>$290 (8,24) & $>$83 (3)   & 4, 5, 6, 10 \\
MWC~342, v1972~Cyg & 20 23 03.61 & +39 29 49.9 & 10.6 & 5.8 & B1 (27) & 1000 (27) & 33000$\pm$15000 (3)  & 4, 5, 27, 28, 29 \\
MWC~361, HD~200775 & 21 01 36.91 & +68 09 47.7 & 7.4 & 5.5 & B3 (11) + B? (30) &  430$^{+160}_{-90}$ (8,24)  & 6000$\pm$2000 (3)  & 4, 5, 6, 14  \\
MWC~1080, v628~Cas & 23 17 25.59 & +60 50 43.6 & 11.6 & 6.0 & B1 (31) & 2200 (32) & 49000$\pm$21000 (3)   & 4, 5, 6, 10, 14 \\
\enddata 
\tablenotetext{a}{Many of the targets are variable stars and these
  magnitudes (V band from Simbad, and H band from 2MASS) are merely
  representative.}  \tablenotetext{b}{Peculiar B[e] star with
  uncertain classification} \tablecomments{References: (1)
  \citet{mora2001}, (2) \citet{wichmann1998}, (3) SED fitting, this
  work, (4) 2MASS; \citet{cutri2003}, (5) Tycho-2; \citet{tycho2}, (6)
  USNO-B Catalog; \citet{usno2003}, (7) \citet{skrutskie1996}, (8)
  \citet{vda1998}, (9) \citet{malfait1998}, (10) \citet{eisner2004},
  (11) \citet{hernandez2004}, (12) \citet{bertout1999}, (13)
  \citet{herbig1988}, (14) \citet{hillenbrand1992}, (15) Guide Star
  Catalog 2.2; \citet{gsc2001}, (16) \citet{ducati2002}, (17)
  \citet{dewinter1997}, (18) New MDM Observatory photometry from 2004
  November, this work, (19) \citet{finkenzeller1985}, (20) \citet{herbst1982}, 
 (21) \citet{perez2004}, (22) DENIS Database; \citet{the2005}, (23)  \citet{sylvester1996},
(24) \citet{hipparcos}, (25) \citet{drew1997}, (26) \citet{mirosh1999b},
(27) \citet{mirosh1999}, (28) \citet{kharchenko2001}, (29) \citet{bergner1995}, 
(30) \citet{pogodin2004}, (31) \citet{cohen1979}, (32) \citet{grankin1992}
}
\end{deluxetable}

\clearpage

\begin{deluxetable}{llll}
%\footnotesize
\tabletypesize{\scriptsize}
\tablecaption{Observing Log
\label{table_obslog}}
\tablehead{
\colhead{Target} &
  \colhead{Date}     & \colhead{Interferometer}  & \colhead{Calibrator Names}\\
          & \colhead{(UT)} & \colhead{Configuration\tablenotemark{a}}  } \startdata
RY~Tau  & 2004 Dec 14 & A28-B10-C00 & HD~27638, HD 27159, HD 26553 \\
\hline
MWC~480  & 2002 Dec 13 &  A35-B15-C15 & HR~1626 \\
        %   & 2003 Nov 24 &  A35-B15-C10 & HR~1626 \\
\hline
AB~Aur & 2002 Dec 8-14 & A35-B15-C15 & HR~1626, SAO~57504 \\
      & 2002 Dec 16,19 & A35-B05-C10 & HR~1626 \\
       & 2003 Feb 19    &             &    \\
       & 2003 Feb 20,22,23    & A25-B15-C10 & HR~1626 \\
       & 2003 Nov 20,22 & A35-B15-C10 & HR~1626\\
       & 2004 Dec 03 & A35-B15-C00 & HR~1626\\
       & 2004 Dec 13 & A28-B10-C00 & HR~1626\\
\hline
HD~45677 & 2003 Nov 29,30 & A35-B15-C10 & HD~46218\\ 
         & 2004 Mar 26 & & HD~46218\\
         & 2004 Dec 13, 14 & A28-B10-C00 & HD~46218\\
         & 2004 Dec 16 & A28-B05-C10 & HD~46218\\
\hline
MWC~166 & 2003 Nov 29-30 & A35-B15-C10 &  HD~50067,  HD~46128 \\
      & 2004 Dec 10    & A35-B15-C00 & HD~47054, HD~49147 \\
\hline
HD~144432  & 2004 Jun 6-8 & A35-B15-C10 & HD~143766 \\
\hline
MWC~863 & 2003 Jun 17,18  & A35-B15-C10 & HR~6153 \\
         & 2004 Mar 16,23-25 & A35-B15-C10 & HR~6153 \\
\hline
MWC~275 & 2003 Jun 12,21 & A35-B15-C10 &  HR~6704, HD~171236 \\
  & 2004 Mar 26  &  & HR 6704, HD~171236 \\
      &2004 Jun 2,7-8 &  &  HD~157546, HD~170657, HD~174596\\
\hline
MWC~297  & 2003 Jun 12, 23 & A35-B15-C10 & HD 171236, HR~7149, HR~7066 \\
              & 2005 Jun 20, 22, 26     & A35-B15-C10 & HR~7066, HR~7149 \\
              & 2005 Jun 29, 30 & A35-B15-C00 & HR~7066, HD~169268 \\
\hline
MWC~614   & 2003 Jun 16-17 & A35-B15-C10 & HD~182101, HD~184502 \\
\hline
v1295~Aql & 2003 Jun 14,17 & A35-B15-C15 & HD~185209 \\
\hline
MWC~342  & 2003 Jun 13 & A35-B15-C10 & HD~191589, HD 199547 \\
\hline
MWC~361-A & 2003 Jun 21,22 & A35-B15-C10 & HR~7967 \\
            & 2003 Nov 27-29 & A35-B15-C10 & HR 7967, HD 197950 \\
            & 2004 Jun 6 & A35-B15-C10 & HD~193664 \\
            & 2004 Dec 10 & A35-B15-C00 & HR~7967\\
            & 2005 Jun 17,18 & A35-B15-C10 & HD~193664 \\
            & 2005 Jun 29    & A35-B15-C00 & HD~193664 \\

\hline
MWC~1080 & 2003 Nov 30 & A35-B15-C10 & HD~221639 \\
            & 2004 Dec 10 & A35-B15-C00 & HR~8881\\
            & 2004 Dec 13 & A28-B10-C00 & HR~8881, HD~221639 \\

\enddata
\tablenotetext{a}{Configuration refers to the location of telescopes A,B,C on
the NE, SE and NE arm respectively; see \S\ref{section:observations} for more details}
\end{deluxetable}
\clearpage
\begin{deluxetable}{llll}

\tabletypesize{\scriptsize}
\tablecaption{Calibrator Information\label{calibrators}
}
\tablewidth{0pt}
\tablehead{
\colhead{Calibrator} & \colhead{Spectral} & \colhead{Adopted Uniform 
Disk}  &
\colhead{Reference(s)}\\
\colhead{Name} & \colhead{Type} & \colhead{Diameter (mas)} & \colhead{}  }
\startdata
HD~26553 & A4III & 0.37$\pm$0.30 & getCal\tablenotemark{a} \\
HD~27159 & K1III & 0.30$\pm$0.10 & getCal \\
HD~27638 & B9V & 0.32$\pm$0.10 & getCal \\
HD~46218 & A5 & 0.36$\pm$0.60 & getCal \\
HD~47054 & B8 & 0.10$\pm$0.15 & getCal\\
HD~49147 & B9.5 & 0.35$\pm$.03 & getCal \\
HD~50067 & K4III & 1.35$\pm$0.35 & getCal \\
HD~143766 & F7V & 0.30 $\pm$0.05 & getCal\\
HD~157546 & B8V  & 0.22$\pm$0.06 & getCal \\
HD~168415 & K4III & 2.6$\pm$0.3 &  Blackbody fit\\ 
HD~169268 & F6III & 0.28 $\pm$ 0.03 & getCal\\
HD~170657 & K1V & 0.58$\pm$0.10 & getCal\\
HD~171236 & K1III & 1.0$\pm$0.5 & getCal \\
HD~174596 & A3V & 0.2$\pm$0.05 & getCal \\
HD~182101 & F6V & 0.4$\pm$0.1  & getCal \\
HD~184502 & B3III & 0.19$\pm$0.10 & getCal \\
HD~185209 & K3III & 0.7$\pm$0.2 & Blackbody fit\\
HD~191589 & K5III & 1.47$\pm$0.37 & getCal \\
HD~193664 & G0V & 0.22$\pm$0.07 & getCal \\
HD~197950 & A8V & 0.32$\pm$0.06 & getCal \\
HD~199547 & K0III & 0.75$\pm$0.15 & getCal \\
HD~221639 & K1V& 0.51$\pm$0.10 & getCal \\
HR~1626  & K0III & 1.2$\pm$0.7 & getCal \\
HR~6704 & K0III & 1.89$\pm$1.47 & getCal \\
HR~7066 & K0I & 2.17$\pm$1.01 & getCal \\ 
HR~7149 & K2III & 1.69$\pm$0.75 & getCal \\
HR 7967 & G8III & 0.76$\pm$0.30 & getCal\\
HR~8881 & M2III & 2.4$\pm$0.5 & \citet{gvb_pasp} \\
SAO~57504 & K0 & 0.58$\pm$0.36 & getCal \\
 \enddata
\tablenotetext{a}{{\em getCal} is maintained and distributed by the
Michelson Science Center (http://msc.caltech.edu)}
\end{deluxetable}

\clearpage

\begin{deluxetable}{lllll}
\tiny
\tabletypesize{\scriptsize}
\tablecaption{Herbig Ae/Be Disk Properties
\label{diskresults2}
}
\tablewidth{0pt}
\tablehead{
\colhead{Target} & \colhead{Dust Fraction} & \multicolumn{2}{c}{Ring Diameter} &  \\
\colhead{Name} & \colhead{at H-band\tablenotemark{a}}  &  \colhead{mas} & \colhead{AU} & \colhead{Comments}
}
\startdata
RY~Tau   & 0.8$\pm$0.05 &  3.99$\pm$0.35 & 0.57$\pm$0.05 & Poor fit w/ no halo emission \\
         &  &              2.85$\pm$0.24 & 0.40$\pm$0.04 & Good fit w/ 10\% halo emission \\
MWC~480 & 0.54$\pm$0.06 & 3.8$\pm$0.3 & 0.50$\pm$0.04 & \\
AB~Aur    & 0.65$\pm$0.10 & 3.2$\pm$0.3 & 0.46$\pm$0.05 & 7.5$\pm$2.5 \% Halo emission \\
HD~45677  & 0.46$\pm$0.03\tablenotemark{b} & $\sim$15.6 & 15.6 & Strong Asymmetry. 
See Table~\ref{hd45677results} \\
          & 0.62$\pm$0.03\tablenotemark{b} & $\sim$12.9\tablenotemark{c} & 12.9 & Gaussian FWHM \\
MWC~166   & 0.1$\pm$0.1 & 10$\pm$5  & 11.5$\pm$5.8 & Weak IR excess. Nearby companion star. \\
HD~144432 & 0.5$\pm$0.1 & 4.1$\pm$1.0  & 0.57$\pm$0.14 &  \\
MWC~863 & 0.65$\pm$0.15 & 4.7$^{+0.8}_{-0.5}$ & 0.71$^{+0.12}_{-0.08}$ & Poor SED fit \\
MWC~275 & 0.59$\pm$0.06 & 3.3$\pm$0.5  & 0.40$\pm$0.06 & 5.0$\pm$2.5 \% Halo emission \\
MWC~297 & 0.60$\pm$0.05 & 5.6$\pm$0.5 & 1.4$0\pm$0.13 & Possible $\simle$5\% Halo \\ 
MWC~614 & 0.39$\pm$0.03\tablenotemark{d} & $\simge$13 & $\simge$3.1 & Suspect nearby companion \\
v1295~Aql & 0.5$\pm$0.1 & 3.48$\pm$0.40 & 1.01$\pm$0.12 & assuming $d=290$pc\\
MWC~342 & 0.75$\pm$0.10 & 3.1$\pm$0.3 & 3.1$\pm$0.3 & \\
MWC~361-A  & 0.6$\pm$0.1 & 3.47$\pm$0.44 & 1.49$\pm$0.19 & Primary component only \\
MWC~1080 & 0.90$\pm$0.10 &  3.11$\pm$0.21  & 6.84$\pm$0.46 & \\
\enddata
\tablenotetext{a}{Best estimate for fraction of H-band light coming from circumstellar material based
on most recent photometry.  Upper and lower limits are based on
SED fitting to diverse data sets and represent the range of possible values given
historical variability.  }
\tablenotetext{b}{H-band excess derived directly from fitting visibility
data. SED analysis for HD~45677 yields estimate for H-band dust 
fraction 0.65$\pm$0.15.}
\tablenotetext{c}{Full-Width at Half-Maximum (FWHM) result from fitting Gaussian profile instead of ring model.}
\tablenotetext{d}{H-band excess derived directly from fitting visibility
data. SED analysis for MWC~614 yields estimate for H-band dust 
fraction 0.37$\pm$0.10.}

\end{deluxetable}

\begin{deluxetable}{lcccl}
\footnotesize
\tablecaption{Closure Phase Results
\label{cpresults}}
\tablehead{
\colhead{Target}  & \colhead{Maximum Closure} & \colhead{Minimum Fringe} & \colhead{Ring Diameter} \\
  & \colhead{Phase\tablenotemark{a} (deg)} & \colhead{Spacing (mas)} & \colhead{(fringes)}} 
\startdata
RY~Tau    &   0.1$\pm$0.4          &  11.5   & 0.35 \\
MWC~480   &   0.1$\pm$2.5          &   8.9   & 0.43 \\
AB~Aur    &  {\bf -4.1$\pm$0.4 }   &  11.7   & 0.27\\
HD~45677  &  {\bf -26.6$\pm$4.0}   &  11.4   & 1.37 \\
MWC~166   &   1.8$\pm$0.9          &  12.1   & 0.83 \\
HD~144432 &  -1.6$\pm$1.3          &  13.5   & 0.30 \\
MWC~863   &   5.1$\pm$2.6          &  11.6   & 0.41  \\
MWC~275   &  -0.6$\pm$0.4          &  11.8   & 0.28 \\
MWC~297   &  {\bf -1.9$\pm$0.4}    &   9.6   & 0.58 \\
MWC~614   &  {\bf 4.1$\pm$1.7}     &   9.3   & $>$1.4 \\
v1295~Aql &   1.9$\pm$1.7          &   9.4   & 0.37  \\
MWC~342   &  -1.8$\pm$1.6          &   8.9   & 0.35 \\
MWC~361-A  &  {\bf 31.0$\pm$1.5}    &  10.2   &  binary  \\
MWC~1080  &  {\bf -1.3$\pm$0.5}    &   9.8   & 0.32 \\
\enddata
\tablenotetext{a}{This column contains the most statistically-significant 
closure phase datum showing a deviation from zero, based on uv-averaged
data. The value is bolded if $>$2-$\sigma$ detection.}
\end{deluxetable}

\clearpage
\begin{deluxetable}{lllllllll}
\tabletypesize{\tiny}
\rotate
\tablecaption{Elongated Ring Models for HD~45677
\label{hd45677results}
}
\tablewidth{0pt}
\tablehead{
\colhead{Model} & \colhead{Dust Fraction} & \colhead{Star Fraction} & \multicolumn{3}{c}{Ring Diameter} & \colhead{Skew} & \colhead{Skew-PA} \\
\colhead{Type} &  \multicolumn{2}{c}{at H-band}  &  \colhead{Major} 
 & \colhead{Minor} & \colhead{PA} & & & 
}
\startdata
Elliptical Ring        & 0.46$\pm$0.01 & 0.54$\pm$0.01 & 17.8$\pm$1.3 & 12.9$\pm$1.4 & 70$\pm$16 & 0 & -- \\
Skewed Elliptical Ring & 0.46$\pm$0.01 & 0.54$\pm$0.01 & 18.5$\pm$1.5 & 13.4$\pm$0.6
 & 77$\pm$13 & 0.92$\pm$0.08 & -31$\pm$8 \\
\enddata
\end{deluxetable}

\clearpage
\begin{deluxetable}{lcccc}
\tablecaption{Astrometric Binary Solutions for MWC~361-A
\label{binarypositions}
}
\tablewidth{0pt}
\tablehead{
\colhead{Epoch}       &  \colhead{$\Delta\alpha$} & \colhead{$\Delta\delta$ }  &
  \colhead{Separation} &\colhead{PA}\\
\colhead{(U.T.)} & \colhead{(mas)} & \colhead{(mas)} & \colhead{(mas)} & \colhead{($\arcdeg$)}  }
\startdata
1998 Jun 13-21 & 0.9$\pm$3.0  & 17.9$\pm$2.6  & 17.9 &   2.7 \\
1998 Sep 29    & 2.9$\pm$3.1  & 17.7$\pm$1.1  & 18.0 &   9.4 \\
2003 Jun 21-22 & 11.4$\pm$4.6 &  3.5$\pm$2.9  & 11.9 &  73.0 \\
2003 Nov 27-29 &  5.8$\pm$3.8 & -5.8$\pm$1.0  &  8.2 & 134.8 \\
2004 Jun 6     &  1.7$\pm$2.6 & -11.3$\pm$1.2 & 11.4 & 171.6 \\
2004 Dec 10    & -3.3$\pm$3.4 & -1.9$\pm$0.6  &  3.9 & 239.8 \\
2005 Jun 17-29 & -2.6$\pm$2.0 & 12.5$\pm$0.8  & 12.8 & 348.3 \\
\hline
\multicolumn{5}{l}{Other binary information:}\\
\multicolumn{5}{c}{H band ratio: 6.5$\pm$0.5, Primary UD diameter: 3.6$\pm$0.5~mas}\\
\hline
\enddata
\end{deluxetable}
\clearpage

\clearpage
\begin{deluxetable}{lcc}

%\tabletypesize{\scriptsize}
\tablecaption{Preliminary Orbital Parameters for MWC~361-A
\label{binaryparams}
}
\tablewidth{0pt}
\tablehead{
\colhead{Parameter} & \colhead{Radial Velocity} & \colhead{Interferometry} \\
 & \colhead{Pogodin et al. 2004} & \colhead{This Work} 
}
\startdata
Period (days) & 1341$\pm$41 & 1377$\pm$25 \\ 
T$_0$ (JD)      & 2449149$\pm$87 & 2449152$\pm$90 \\
a$_1$~sin(i) (AU) & 1.33$\pm$0.08 & --- \\
a (mas)       & --- & 15.14$\pm$0.70 \\
e             & 0.29$\pm$0.07 & 0.30$\pm$0.06 \\
i ($\arcdeg$)  & --- & 65$\pm$8 \\
$\omega$ ($\arcdeg$) & 203$\pm$22 & 224$\pm$16 \\
$\Omega$ ($\arcdeg$) & --- & -0.2$\pm$7.6 \\
K (km/s)      & 11.2$\pm$0.7 & ---\\
f(M$_2$) (\msun)      & 0.175$\pm$0.035 & --- \\
$M_1+M_2$ (\msun)\tablenotemark{a} &  \multicolumn{2}{c}{$10.4^{+20.5}_{-5.9}$ }
\enddata
\tablenotetext{a}{For distance d$=430^{+160}_{-90}$~pc; see description of system mass 
estimate in \S\ref{mwc361orbit}.}
\end{deluxetable}
\clearpage

%FIGURE 1
% fig_uvcov
% Big Multi-plot set of figures which show all the UV Coverage.

%FIGURE 2
% fig_allvis2
% Big plot showing Visibility2 of all DATA

%FIGURE 3
% fig_allcp
% Big plot showing all cp data.

\clearpage

\clearpage

\clearpage
\clearpage

\clearpage

\clearpage

\clearpage

\clearpage

\clearpage

\clearpage

\clearpage

\clearpage

\clearpage

\clearpage

\clearpage

\clearpage

\clearpage

\clearpage

\clearpage

\clearpage

%% Deluxetable environment provided by the AASTeX package or the LaTeX
%% table environment.  Use of deluxetable is preferred.
%%

%% Three table samples follow, two marked up in the deluxetable environment,
%% one marked up as a LaTeX table.

%% In this first example, note that the \tabletypesize{}
%% command has been used to reduce the font size of the table.
%% Note also that the \label command needs to be placed 

\end{document}